\documentclass[12pt]{article}
\usepackage{jheppub}

\usepackage{amsfonts,amsopn,amsmath,amssymb}
\usepackage{amsthm}
\def\beq{\begin{equation}}  
\def\eeq{\end{equation}}  
\def\bea{\begin{eqnarray}}  
\def\eea{\end{eqnarray}}  
\def\half{\frac{1}{2}}  
 
\def\Tr{{\rm Tr}}

\def\inn{\,\in\,}
\def\erfc{\text{erfc}}

\newcommand{\ZZ}{{\mathbb Z}}%Integers
\newcommand{\CL}{\mathcal{L}}

\newcommand{\CN}{\mathcal{N}}

\newcommand{\CQ}{\mathcal{Q}}

\newcommand{\CD}{\mathcal{D}}
\newcommand{\CH}{\mathcal{H}}

\renewcommand{\Im}{{\rm Im}}
\renewcommand{\Re}{{\rm Re}}

\newcommand{\sgn}{\mbox{sgn}}

\newcommand{\IR}{\mathbb{R}}
\newcommand{\IC}{\mathbb{C}}
\newcommand{\IZ}{\mathbb{Z}}
\newcommand{\IH}{\mathbb{H}}

\def\Z{\mathbb{Z}}

\newcommand{\ndt}{\noindent}

\def\e{\epsilon}

\def\bea{\begin{eqnarray}}
\def\eea{\end{eqnarray}}
\def\be{\begin{equation}}
\def\ee{\end{equation}}
\def\ba{\begin{align}}
\def\ea{\end{align}}
\def\bse{\begin{subequations}}
\def\ese{\end{subequations}}

\newcommand{\bem}{\begin{pmatrix}}
\newcommand{\eem}{\end{pmatrix}}

\def\qb {\bar{q}}

\def\={\;  = \;}
\def\+{\, + \,}

\def\wt{\widetilde}
\def\wh{\widehat}
\def\bar{\overline}

\def\rt2{\sqrt{2}}

\renewcommand{\Im}{\mbox{Im}}
\renewcommand{\Re}{\mbox{Re}}

\def\vth{\vartheta}
\def\ve{\varepsilon}
\def\v{\varphi}

\def\s{\sigma}

\def\t{\tau}
\def\a{\alpha}
\def\b{\beta}
\def\d{\delta}

\def\l{{\lambda}}

\def\G{\Gamma}

\def\AA{\mathcal{A}}

\def\AP#1{\mathcal A_{#1}}

\def\mypmod#1{\; ({\rm mod} \; {#1})}

\def\p{\partial}
\def\tbar{\bar \tau}

\def\ubar{\bar u}

\newcommand{\dd}{{\rm d}}

\newcommand{\bPhi}{\overline{\Phi}}
\newcommand{\bP}{\overline{P}}
\newcommand{\bp}{\overline{p}}
\newcommand{\bphi}{\overline{\phi}}
\newcommand{\bpsi}{\overline{\psi}}
\newcommand{\bQ}{\overline{Q}}
\newcommand{\bF}{\overline{F}}

\newcommand{\bD}{\overline{D}}
\newcommand{\bchi}{\overline{\chi}}
\newcommand{\blambda}{\overline{\lambda}}
\newcommand{\bsigma}{\overline{\sigma}}

\title{A holomorphic anomaly in the elliptic genus}

\author{Sameer Murthy}
\affiliation{Department of Mathematics, King's College London \\
The Strand, London WC2R 2LS, U.K.}
\abstract{We consider a class of gauged linear sigma models (GLSMs) in two dimensions that flow 
to non-compact $(2,2)$ superconformal field theories in the infra-red, a prototype of which is 
the~$SL(2,\IR)/U(1)$ (cigar) coset. 
We compute the elliptic genus of the GLSMs as a path-integral on the torus using supersymmetric localization.
We find that the result is a Jacobi-like form that is non-holomorphic in the modular parameter~$\t$ of 
the torus, with mock modular behavior. This agrees with a previously-computed expression 
in the cigar coset. 
We show that the lack of holomorphicity of the elliptic genus arises from the contributions 
of a compact boson carrying momentum and winding excitations. This boson has an 
axionic shift symmetry and plays the role of a compensator field that is needed to cancel the 
chiral anomaly in the rest of the theory.
}
\keywords{GLSM, chiral anomaly, mock modular forms}

\begin{document}

\maketitle

\section{Introduction and summary}

The setting of this paper is two-dimensional quantum field theories with~$(2,2)$ supersymmetry with left 
and right-moving~$U(1)$ R-symmetry. 
An interesting quantity to consider in such theories is the partition function on the torus. 
If we give the fermions periodic boundary conditions, and switch on a chemical potential~$z$ 
for the the left moving R-charge, the resulting partition function~$\chi(\t,z)$, called 
the \emph{elliptic genus}, is a function of the modular parameter~$\t \in \IH$ of the torus and the chemical 
potential~$z \in \IC$ 
\cite{Schellekens:1986yi, Schellekens:1986xh, Pilch:1986en, Witten:1986bf, Witten:1987cg, Alvarez:1987wg,Alvarez:1987de}. 
In the Hamiltonian formalism,
\be \label{defell}
\chi(\t,z) \= \Tr_{\CH_{RR}} \, (-1)^F q^{L_0} \,  \qb^{\overline L_0} \, \zeta^{J_0} \= \langle 1 \rangle_{(z,+)} \, ,
\ee
where~$\CH_{RR}$ is the Hilbert space of the theory with periodic $(++)$ boundary conditions for the 
fermions,~$L_{0}$ and~${\overline L_{0}}$ are the left and right-moving Hamiltonians,~$J_{0}$ is the left-moving 
R-charge,~$F$ is the fermion number operator, and we have defined~$q=e^{2 \pi i \tau}$, $\zeta = e^{2 \pi i z}$.

The elliptic genus can equivalently be thought of as the functional integral of the theory with periodic boundary 
conditions on the right-moving fermions and twisted boundary conditions on the left-moving fermions with twist 
parameter~$z$, as we have indicated on the right-hand side of~\eqref{defell}. 
The twisted boundary conditions can be implemented by turning on a constant background gauge field. 
The invariance of the functional integral under coordinate transformations of the torus of the 
type~$\t \to \t+1$, $\t \to -1/\t$ implies that the elliptic genus enjoys modular transformation properties. 
Similarly, the symmetry of the quantum field theory under large gauge transformations implies nice transformation 
properties under~$z \to z+1$ and $z \to z+\t$. 
These properties can be summarized concisely in the statement that the elliptic genus is a Jacobi form 
of weight zero\footnote{In the context of the geometric 
definition of elliptic genus, this was proven in~\cite{Zagier:Ellgen}.}. 

As was explained in~\cite{Witten:1986bf}, the elliptic genus is invariant under continuous changes 
of the parameters of the Hamiltonian. The basic argument is that massive representations of the 
right-moving supersymmetry algebra come in pairs with the same value of energy ${\overline L_{0}}\ge 0$ and 
opposite values of~$(-1)^{F}$, and therefore do not contribute to the elliptic genus. 
A continuous change of parameters in the supersymmetric Hamiltonian will only affect the massive 
representations, and will therefore not affect the elliptic genus. 
The same argument implies that the elliptic genus is a \emph{holomorphic} function of~$\t$, as the only 
right-movers that contribute are in the ground state with~${\overline L_{0}}=0$. 

The above heuristic argument can be made more precise with the assumption of a discrete spectrum as 
in the case of compact target spaces, or, more generally, in a rational superconformal field theory (SCFT).
On the other hand, when there is a continuum in the spectrum, then the trace in~\eqref{defell} needs to be 
well-defined, and the natural language to describe such a situation is that of density of states. 
Some explicit examples of interesting non-compact SCFTs have been known for 
quite a while. The basic example with all the non-trivial features is the~$SL(2,\IR)/U(1)$ 
SCFT\footnote{This is similar to how Liouville theory can be thought of as the basic example for 
bosonic non-compact models. This parallel goes further because the supersymmetric cigar theory is 
mirror-symmetric to~$\CN=2$ Liouville theory~\cite{Hori:2001ax}. From another point of view, 
the~$SL(2,\IR)/U(1)$ theory can be thought of as the analytic continuation of~$SU(2)/U(1)$ theories that 
are the~$\CN=2$ minimal models.} popularly known as the ``cigar'' theory due to the semi-infinite shape 
of the target space manifold. 

The elliptic genus of the~$\CN=(2,2)$ supersymmetric~$SL(2,\IR)/U(1)$ theory was recently computed 
in~\cite{Troost:2010ud,Eguchi:2010cb,Ashok:2011cy}, by explicitly solving the path-integral of the theory. 
The result is interesting because the cigar elliptic genus~$\chi^{\rm cig}(\t,z)$ is  \emph{not holomorphic} 
in~$\t$. The function~$\chi^{\rm cig}(\t,z)$ is a product of a usual Jacobi form and a 
\emph{completed Appell-Lerch sum}~\cite{Zwegers:2002, Zagier:2007}. (See~\eqref{ALcig} and 
Appendix~\S\ref{JacApp} for the precise expression.)
Appell-Lerch sums are intimately related to a very interesting class of functions called mock-modular 
forms~\cite{Zwegers:2002, Zagier:2007, DMZ}.
The key feature of a function~$\wh f(\t)$ of this type is that it transforms like a holomorphic modular form 
of weight~$k$, but it suffers from a \emph{holomorphic anomaly}:
\be \label{mmf}
(4 \pi \t_{2})^{k} \, \p_{\tbar} \wh f(\t) = -2 \pi i \, \overline{g(\t)} \, ,
\ee
where~$g(\t)$ is a holomorphic modular form called the \emph{shadow} of~$f$ with weight~$2-k$. 
The cigar elliptic genus~$\chi^{\rm cig}(\t,z)$ shares this feature as exhibited in Equation~\eqref{holanomchi}.  

The non-holomorphicity of the elliptic genus of the cigar theory has been attributed to the difference in 
the density of states between the bosons and fermions in the spectrum of normalizable states~\cite{Ashok:2011cy}. 
This difference is dictated by the ratio of the reflection coefficients in the cigar coset 
theory\footnote{If one considers the limit in which the 
tip of the cigar is pushed away infinitely to get a (singular) pure linear-dilaton theory, 
%%the reflection coefficient vanishes and 
the elliptic genus collapses to a holomorphic expression as for free field theories.}.

\vspace{0.1cm}

\ndt The computation of the cigar elliptic genus leads to some natural questions:
\begin{enumerate}
\vspace{-0.2cm}
\item Is there a simple setting to understand the essential features of the non-compact models 
without having to understand all the details of strongly coupled CFTs?

\vspace{-0.2cm}
\item Can the cigar result be generalized to a larger class of models? 

\vspace{-0.2cm}
\item Can the holomorphic anomaly be understood as arising from an anomaly in a physical symmetry? 

\end{enumerate}
\vspace{-0.2cm}

In this paper, we shall address and answer these questions in the context of gauged linear sigma models 
(GLSMs)~\cite{Witten:1993yc}.
In particular, we study examples of two-dimensional~$(2,2)$ supersymmetric quantum field theories with 
one~$U(1)$ gauge field multiplet and a set of chiral multiplets with charges~$Q_{i}$ under the gauge symmetry, 
that flow to non-compact theories. In two dimensions, the gauge coupling is super-renormalizable and so these 
theories undergo an RG flow. Their infra-red fixed points are generically interacting~$(2,2)$ SCFTs. By the 
same arguments as above, we expect that the elliptic genus does not depend on the energy scale.

The GLSM that flows to the cigar coset was introduced and studied in an impressive paper by Hori and 
Kapustin~\cite{Hori:2001ax}, and was later generalized to a class of models by the same authors 
in~\cite{Hori:2002cd} -- these latter theories flow to SCFTs that are not explicitly known, and are conjectured 
to arise on NS5-branes wrapped on various curved surfaces\footnote{There has been a very recent 
conjecture~\cite{Ashok:2013zka} for the elliptic genus of these theories.}. 
The models are known to have a radial non-compact 
direction in the IR, with a compact surface fibered over this direction. 

In this paper, we evaluate the elliptic genus of the class of GLSMs introduced in~\cite{Hori:2001ax,Hori:2002cd} using 
supersymmetric localization, by adapting the method developed in~\cite{Benini:2013nda} for the compact models. 
We find that the elliptic genus has a simple expression in terms 
of a two-dimensional integral over the Wilson lines around the torus of the~$U(1)$ gauge field. This expression has a 
holomorphic anomaly in~$\t$, and, in the simplest case (Equation~\eqref{chifinalcig}), it is equal to the elliptic genus 
of the cigar theory as computed 
in~\cite{Troost:2010ud, Eguchi:2010cb,Ashok:2011cy}. In the other cases, it is equal to the expression conjectured 
in~\cite{Ashok:2013zka}. 

These non-compact models are characterized by an anomaly in the chiral rotation of the fermions due to the fact 
that~$\sum_{i} Q_{i} \neq 0$. The models, however, do have conserved currents that rotate the fermions chirally 
because there is an additional bosonic compensator field with an axionic shift symmetry. On adding its derivative 
to the chiral fermion current, the anomaly is cancelled. This compensator field, and the corresponding 
superfield~$P$, is at the heart of many of the interesting features of these models.

From our ultra-violet analysis, we show that the non-holomorphicity in the elliptic genus 
arises from the contributions of the compact boson~$\Im \, p$, the imaginary part of the 
lowest component of the chiral superfield~$P$. This is in spirit similar to the identification of the holomorphic anomaly 
in the topological string as coming from the zero mode at the boundary of moduli space\footnote{Indeed a baby example 
of a mock modular form is the quasi-holomorphic modular form~$\wh E_{2}(\t) = E_{2}(\t)-\frac{3}{\pi \t_{2}}$ that arises 
in the topological string.}. 
In our case, the zero mode is compact, and instead of a power of~$\t_{2} \equiv \Im \, \t$, we get a full theta function of 
the compact (non-chiral) boson~$\Im \, p$. We have thus identified a simple physical source of the holomorphic anomaly 
in such models, namely as arising from the chiral anomaly of a two-dimensional field theory. This seems to be 
an example of the general idea that introducing compensators for anomalous symmetries destroys some other nice 
property of the theory. See~\cite{deWit:1985bn} for a nice discussion and examples of this phenomenon in various contexts.

The plan of the paper is as follows. In \S\ref{GLSMCig}, we review the prototype GLSM of Hori and Kapustin 
that flows to the supersymmetric~$SL(2,\IR)/U(1)$ SCFT. We discuss its symmetries, and discuss how the chiral 
anomaly manifests itself in the UV and the IR theory. We then review the result 
of~\cite{Troost:2010ud} for the elliptic genus of the cigar SCFT. In~\S\ref{ellgenGLSM}, we review the method 
of~\cite{Benini:2013nda} to compute the elliptic genus of compact GLSMs using localization. 
We then adapt this method to the non-compact models of interest to us, derive their 
elliptic genera, and show how the non-holomorphic contributions can be understood as arising  
from the contributions of the compensator multiplet~$P$. In \S\ref{holanom}, we express the holomorphic 
anomaly of all the theories as a contour integral. In \S\ref{discuss}, we end with some comments, a 
discussion of issues that would be interesting to resolve, and by sketching some directions for future research. 

\ndt \emph{Note added}:  
While this paper was being prepared for publication, the author received communication of~\cite{Ashok:2013pya} 
which contains overlapping results. 

\section{The RG flow from a GLSM to the cigar \label{GLSMCig}}

The prototype GLSM that we shall study is the one introduced by Hori and Kapustin~\cite{Hori:2001ax}, 
which we now review. The field content consists of a vector superfield~$V$ with 
components~$(v_{\mu}, \sigma , \bsigma, \lambda_{\pm} , \blambda_{\pm} , D)$ in the Wess-Zumino gauge, 
a chiral superfield $\Phi$ with components $(\phi, \bphi , \psi_{\pm} , \bpsi_{\pm} , F, \bF)$ and a chiral 
superfield~$P$ with components~$(p, \bp , \chi_{\pm} , \bchi_{\pm}  , F_{P}, \bF_{P})$. 
We follow the conventions of~\cite{Witten:1993yc} in which the $\pm$ subscript implies left and right moving fields. 
The fields~$v_{\mu}$ and~$D$ are real while the other fields are complex with the bar denoting complex conjugation. 
For reasons that will soon become clear, we shall refer to the~$P$-superfield as the compensator superfield.

Under the~$U(1)$ gauge transformation
\be \label{gge}
V \to V - i \Lambda + i \overline{\Lambda} \,  \qquad \text{with}  \quad \overline{D}_{+} \Lambda = \overline{D}_{-} \Lambda =0 \, ,
\ee
the chiral superfields transform as follows:
\be \label{ggetrans}
\Phi \to e^{i \Lambda} \Phi \, , \qquad P \to P + i \Lambda \, .
\ee
The field~$\Im \, {P}$ is periodically identified with period~$2 \pi$. 
The above inhomogeneous transformation of the superfield field~$P$ is at the root of many of the 
interesting features of this system. 

The action is\footnote{In the context of string theory, we choose conventions for the fundamental length scale in target space 
in which~$\ell_{s}^{2}\equiv \a'=2$.}
\be \label{action}
S={1\over 4\pi}\int\dd^2x\,\dd^4\theta \, \Big[ \bPhi \, {\rm e}^V\Phi+{k\over 4}(P+\bP+V)^2-{1\over 2e^2}|\Sigma|^2 \Big] \, , 
\ee
where $\Sigma=\bD_+ D_-V$ is a twisted chiral superfield, obeying $\bD_+\Sigma=D_-\Sigma=0$. 
The action in terms of the component fields is:
\bea \label{actioncomp}
S = &&\!\!\!\!{1\over 4\pi}\int\dd^2x~\Biggl[
-\CD^{\mu}\bphi\CD_{\mu}\phi
+i\bpsi_-(\CD_0+\CD_1)\psi_-
+i\bpsi_+(\CD_0-\CD_1)\psi_+
+D|\phi|^2+|F|^2~~~~~~~~~~~
\nonumber\\[-0.1cm]
&&~~~~~~~~~~~~~~~
-|\sigma|^2|\phi|^2
-\bpsi_-\sigma\psi_+
-\bpsi_+\bsigma\psi_-
-i\bphi\lambda_-\psi_+
+i\bphi\lambda_+\psi_-
+i\bpsi_+\blambda_-\phi
-i\bpsi_-\blambda_+\phi
\nonumber\\[0.1cm]
&&~~~~~~~~~~~~
+{k\over 2}\Bigl(-\CD^{\mu}\bp\CD_{\mu}p
+i\bchi_-(\partial_0+\partial_1)\chi_-
+i\bchi_+(\partial_0-\partial_1)\chi_+
+D(p+\bp)+|F_P|^2
\nonumber\\
&&~~~~~~~~~~~~~~~
-|\sigma|^2
+i\chi_+\lambda_-
-i\chi_-\lambda_+
+i\bchi_+\blambda_-
-i\bchi_-\blambda_+\Bigr)
\nonumber\\
&&~~~~~~~~~~~~
+{1\over 2e^2}\Bigl(
-\partial^{\mu}\bsigma\partial_{\mu}\sigma
+i\blambda_-(\partial_0+\partial_1)\lambda_-
+i\blambda_+(\partial_0-\partial_1)\lambda_+
+F_{01}^2+D^2\Bigr)\Biggr].
\eea
In the above expression, $\CD_{\mu}\phi = (\p_{\mu} + i v_{\mu})\phi$ and $\CD_{\mu}\psi_{\pm} = (\p_{\mu} + i v_{\mu})\psi_{\pm} $ 
are the standard covariant derivatives, while $\CD_{\mu}p =\partial_{\mu}p+iv_{\mu}$. 
Note that the fermions~$\chi_{\pm}$ do not couple to the gauge field. One can add a Fayet-Iliopoulos term 
to the above system, but this can be absorbed into P, and a theta angle is not included because one wants 
to preserve worldsheet parity.

The theory~\eqref{action} is free in the ultra-violet, and is super-renormalizable. 
The mass of the gauge field and its superpartners is set by the scale~$e\sqrt{2k}$, below which one can 
integrate out the vector multiplet and set the D-term to zero. These steps, along with a gauge fixing condition, 
allow us to solve for all the other fields in terms of~$\Phi$ and obtain a target-space metric~\cite{Hori:2001ax}. 
Defining the variables~$u= {\rm arcsinh}\big(\sqrt{\frac2k}|\phi| \big)$, $\psi=\arg \phi$ with~$\psi\sim \psi + 2 \pi$, 
the target-space metric is:
\be\label{cigarlike}
ds^2=2k\left(\cosh^4 u \ du^2+\tanh^2 u\ d\psi^2\right) \, . 
\ee

The metric \eqref{cigarlike} is smooth near the origin~$u=0$, and as~$u \to \infty$ it approaches a flat metric on a cylinder. 
Topologically speaking, this manifold has the shape of a semi-infinite cigar, but the metric is not the target space of a conformal field theory 
and it undergoes a further RG flow. As explained in~\cite{Hori:2001ax}, the end-point of the flow is the~$\frac{SL(2,\IR)_{k}}{U(1)}$ 
SCFT with central charge $c=3+\frac6k$. In the large $k$ limit, the coset has a geometric picture as a sigma model with the 
metric: 
\be\label{cigarmetric}
ds^2=2k\left(du^2+\tanh^2 u\ d\psi^2\right).
\ee
whose curvature is proportional to $1/k$. In addition, there is a non-trivial background dilaton:
\be \label{dilprof}
\Phi_{\rm dil} = \Phi_{{\rm dil} \, 0} - \log \cosh u \, . 
\ee
The target-space metric and dilation obey the equation $2D_aD_b\Phi_{\rm dil} + R_{ab}=0$, 
where $D_a$ is the covariant derivative and $R_{ab}$ is the curvature in target-space~\cite{Witten:1991yr}.

Asymptotically as $u \to \infty$, the cigar theory consists of a linear dilaton direction $\rho=\sqrt{2k} \,u$ with 
slope $Q=\sqrt{\frac{2}{k}}$, and a $U(1)$ direction $\theta=\sqrt{2k} \, \psi$ with $ \theta \sim \theta +2 \pi \sqrt{2k}$, 
and two fermions $(\psi_{\rho}, \psi_{\theta})$. 
Together, they make up an $N=2$ SCFT with the following holomorphic currents of almost free fields (see e.g.~\cite{Murthy:2003es}):
\bea\label{ntwowss}
 T_{\rm cig} & \= & -\half (\partial \rho)^2 -\half (\partial\theta)^2 - \half
 (\psi_\rho\partial \psi_\rho + \psi_\theta \partial \psi_\theta)
 - \half Q \partial^2\rho \, , \cr
 G^\pm_{\rm cig} & \= &\frac{i}{2} (\psi_\rho \pm i\psi_\theta)\partial(\rho \mp
 i\theta) +\frac{i}{2} Q\partial (\psi_\rho \pm i\psi_\theta) \, , \cr
 J_{\rm cig} & \= & -i\psi_\rho\psi_\theta +iQ\partial \theta \, ,
\eea
as well as their anti-holomorphic counterparts. 

The structure of the superconformal currents in the full~$\frac{SL(2,\IR)_{k}}{U(1)}$ Kazama-Suzuki coset theory 
are more complicated than in the almost-free asymptotic region~\eqref{ntwowss}.
One can, nevertheless, solve the model using the algebraic approach which is exact in~$k$.

\subsection{Anomalous and conserved symmetries}

At the classical level, the theory~\eqref{action} has both vector and axial $U(1)$ R-symmetries, under which 
the lowest components of the superfields $\Phi,P$ and $\Sigma$ have 
charges~$(q_V,q_A)=(0,0),(0,0)$ and $(0,2),$ respectively. Equivalently, we can discuss the linear 
combinations that correspond to the left- and right-moving R-symmetries with currents~$j_{L,R} = {1\over 2}(j_A\pm j_V)$.
The components of the right-moving R-current $j_R^{\pm}$ is:
\bea 
&&j_R^+=\psi_-\bpsi_-+{k\over 2}\chi_-\bchi_-
+{i\over 2e^2}\left(\partial_-\bsigma\sigma-\bsigma\partial_-\sigma\right),
\label{jR}
\\
&&j_R^-={1\over 2e^2}\blambda_+\lambda_+
+{i\over 2e^2}\left(\partial_+\bsigma\sigma-\bsigma\partial_+\sigma\right).
\eea
In the IR limit $e^2\to\infty$ where the $\Sigma$ multiplet
becomes very massive, $j_R^-$ vanishes and $j_R^+$
obeys the right-moving condition $\partial_+j_R^+=0$ classically.

At the quantum level this condition is violated, and one has a chiral anomaly~\cite{Hori:2001ax}:
\beq
\partial_{\mu}j_R^{\mu}= 2 F_{+-} \, ,
\label{Aanomaly}
\eeq
with $F_{+-} =\partial_+v_--\partial_-v_+$.
In general, one can modify the right-moving chiral current as
\be
(j_R^+\,,\, j_R^-) \to (j_R^+-v_-\,,\, j_R^-+v_+) \, ,
\ee
to get a current which is conserved, but the new current is not gauge-invariant. 

The interesting feature of the GLSM~\eqref{action} is that the anomalous chiral 
current can be modified in a \emph{gauge-invariant} way so that it is conserved. 
The reason this is possible is the presence of the field ${\rm Im}\,p$ that shifts 
like an axion under the gauge symmetry. In terms of the gauge-invariant field
\beq
A_{\mu}=\partial_{\mu}{\rm Im}\,p+v_{\mu} \, , 
\eeq
we have $F_{+-} =\partial_+A_--\partial_-A_+$.
The modified axial current
\be
\widetilde{j}_R^+=j_R^+-2 A_-,\quad \widetilde{j}_R^-=j_R^-+2A_+,
\ee
is then gauge-invariant and conserved.

This can be lifted to the ultra-violet theory. 
Using the explicit expressions of the current~$j_{R}$~\eqref{jR} and the equation of motion of ${\rm Im}\,p$, one finds
\beq \label{anomjR}
\partial_+\left(\psi_-\bpsi_-+{k\over 2}\chi_-\bchi_-
+{i\over e^2}\sigma\partial_-\bsigma\right)
+\partial_-\left({1\over 2e^2}\blambda_+\lambda_+
-{i\over e^2}\bsigma\partial_+\sigma\right)=F_{+-}=2\partial_+A_- \, . 
\eeq
Using the~$\CQ$-exact expression:
\be \label{Qexcur}
\blambda_+\lambda_+-2i\bsigma\partial_+\sigma =\{\bQ_+,\bsigma\blambda_+\} \, ,
\ee
and a similar complex conjugate expression, we find 
\beq \label{concur}
\partial_+\left(\psi_-\bpsi_-+{k\over 2}\chi_-\bchi_-
+{i\over e^2}\sigma\partial_-\bsigma-2A_-\right)=0~~
\mbox{modulo $\{\bQ_+,\cdots\}$} \, .
\eeq
These expressions have a supersymmetric generalization. The current~$j_{R}$ is the bottom component of 
a current superfield which has an anomaly that is the $(1+1)$-dimensional version of the Konishi anomaly~\cite{Konishi:1983hf}. 
One can build a modified superfield that obeys a supersymmetric chirality condition whose bottom component is 
equation~\eqref{concur}. The~$\CQ$-exact expression~\eqref{Qexcur} will be useful for us later in~\S\ref{ellgenGLSM} as well.

From the cigar point of view, the non-conservation of the chiral rotation  can be understood as due to the anomaly 
at one-loop in the $U(1)$ current $j_R$ which rotates only the right-moving fermions, that is caused by the curvature of the cigar:
\be\label{anomalycig}
 \p_\alpha j_R^\alpha = R (\epsilon^{\alpha\beta}\,\p_\alpha\rho\,\p_\beta\theta) ,
\ee
where $R = -2 D^a D_a \Phi_{\rm dil} = {-Q^2 \over 2\cosh^2{Q\rho\over 2}}$ is the Ricci curvature of the cigar~\cite{Murthy:2003es}. 

Due to the special form of the curvature in two dimensions, we can define a new current which {\it is} conserved. 
Changing to complex coordinates on the worldsheet, this current is the sum of the chiral rotation and another piece 
proportional to the left moving momentum:
\be\label{conscurr}
 \overline{\p} j_G :=  \overline{\p} (j_R + Q (\tanh{Q\rho\over 2}) \p \theta) = 0
\ee
which reduces to the $U(1)$ $R$ current of the  $\CN=2$ SCFT \eqref{ntwowss} in the asymptotic region. 

In addition to these R-symmetries, there is also a non-R symmetry whose only effect is to shift the field~$\Im \, p$ 
by a constant. In the infra-red, this becomes the momentum of the cigar theory. The winding around this circle is 
not conserved in the full theory. 

\subsection{The elliptic genus of the cigar SCFT \label{ellgencig}}

The elliptic genus of the $SL(2,\IR)_{k}/U(1)$ SCFT has recently been explicitly 
computed in \cite{Troost:2010ud,Eguchi:2010cb,Ashok:2011cy}\footnote{The holomorphic part of this partition 
function had been presented earlier in \cite{Eguchi:2004yi}.} by evaluating the functional integral for the Euclidean 
version of the cigar, i.e.~the coset~$H_{3}^{+}/U(1)$ with $H_{3}^{+} = SL(2,\IC)/SU(2)$. 
The evaluation is based on the techniques developed 
in~\cite{Gawedzki:1991yu,Gawedzki:1988nj, Karabali:1989dk,Schnitzer:1988qj} to compute the functional integral of~$G/H$ 
WZW cosets. This is 
done by expressing the coset as~$G \times H^{\IC}/H$ where~$H^{\IC}$ is a complexification of the 
subgroup~$H$ that is gauged, and adding a~$(b,c)$ ghost system of central charge~$c=-\dim(H)$. 
The three pieces are coupled only via zero modes. 

In the case of the supersymmetric~$SL(2,\IR)_{k}/U(1)$ coset, there is a 
bosonic~$H_{3}^{+}$ WZW model at level $k+2$ of which a~$U(1)$ subgroup is gauged, and 
two free fermions $\psi^{\pm}$ (and their right-moving counterparts). 
The coset~$H^{\IC}/H$ is represented by a compact boson~$Y$. The zero mode that couples the various pieces 
corresponds to the holonomy of the gauge field around the two cycles of the torus which is represented by a 
complex parameter~$u = s_{1} \t +s_{2}$. 
The bosonic~$SL(2,\IR)$, the two fermions, the~$Y$ boson, and the~$(b,c)$ ghosts are all solvable theories and are 
coupled by this parameter~$u$ that has to be integrated over the torus $E(\t)=\IC/(\IZ \tau + \IZ)$.

We refer the reader to~\cite{Troost:2010ud,Eguchi:2010cb,Ashok:2011cy} for the details of the computation. 
Here we quote the result\footnote{This expression should be taken to hold for real values of~$z$. 
An expression for the cigar elliptic genus has been derived in~\cite{Ashok:2014nua} for arbitrary complex values 
of~$z$ which agrees with the formula~\eqref{ALcig}.} % ZZZ What about Eguchi-Sugawara?
for the elliptic genus of the supersymmetric cigar theory at level~$k$:
\be\label{chicig}
\chi^{\rm cig}(\t,z) \= k  \int_{0}^{1} \dd s_{1} \, \dd s_{2}  \, 
\frac{\vth_1(\t,-z -\frac{z}{k} + s_{1} \t+s_{2}) }{\vth_1(\t,-\frac{z}{k}+s_{1} \t+s_{2})} \, 
\sum_{m,w} \ e^{2 \pi i z w - \frac{\pi k}{\t_{2}} |m + w \t + s_{1} \t+s_{2} |^{2}} \, . 
\ee
The theta function appearing in this expression is the odd Jacobi theta function, we recall its definition in 
Appendix~\S\ref{JacApp}.  

As was shown in~\cite{Troost:2010ud}, the integral~\eqref{chicig} can be rewritten in terms of a function 
called an Appell-Lerch sum. We briefly summarize the definition and some properties of Appell-Lerch sums 
in~\S\ref{JacApp}. These functions are holomorphic in~$\t$ and meromorphic in~$z$, and they do not transform 
nicely under modular transformations. However, one can add to them a simple non-holomorphic function such 
that the new function, called the completed Appell-Lerch sum, is modular. 
As mentioned in the introduction, this property is the defining feature of mock-modular forms, which are 
one-variable functions of~$\t$. A precise relation between the two-variable Appell-Lerch sums and 
one-variable mock-modular forms is spelt out in~\cite{DMZ}.

In terms of the completed Appell-Lerch sums~$\wh \AA_{1,k} (\t,z)$ of weight~1 and index~$k$, 
we have~\cite{Eguchi:2010cb}:
\be \label{ALcig}
\chi^{\rm cig}(\t,z) =  -\frac{1}{k} \frac{\vth_{1}(\t,z)}{\eta(\t)^{3}} \, \sum_{\a,\b\in \IZ/2k\IZ}  
\, e^{2\pi i \frac{\a \b}{k}} \, q^{\frac{\a^{2}}{k}} \, \zeta^{\frac{2\a}{k}} \, 
\wh \AA_{1,k} \Big(\t,\frac{z+\a\t+\b}{k}\Big) \, .
\ee
The overall normalization is such that the Witten index~$\chi^{\rm cig}(\t,0)=1$.

The normalization of the R-charge is such that the elliptic genus transforms as a Jacobi form 
of weight 0 and index $\frac12+\frac1{k}$. Recalling that the central charge of the cigar is~$c=3+\frac6{k}$, 
this is consistent with the normalization used in Calabi-Yau manifolds of complex 
dimension~$d$, the elliptic genus of which is a Jacobi form of index~$d/2$~\cite{Zagier:Ellgen}.

\section{The elliptic genus of the GLSM \label{ellgenGLSM}}

The elliptic genus of GLSMs for compact theories has been computed in the 1990s following~\cite{Witten:1993yc}, 
using the fact that one can compute it in the free-field limit. Very recently, this problem has been revisited 
in~\cite{Benini:2013nda} in the context of a direct evaluation of the path-integral using supersymmetric localization. 
This is a technique that has been applied recently in a variety of circumstances in supersymmetric field theories 
(following the work of~\cite{Pestun:2007rz}) including two-dimensional field 
theories~\cite{Gadde:2013dda, Benini:2012ui, Doroud:2012xw}, as well as in 
supergravities~\cite{Dabholkar:2010uh, Dabholkar:2011ec},  to compute functional integrals that are independent 
of the coupling constants, or in related situations, independent of the energy 
scales~e.g.~\cite{Jockers:2012dk, Gomis:2012wy, Murthy:2013xpa}. 

The problem at hand is to compute the functional integral of a GLSM on a two-dimensional torus 
with periodic boundary conditions on the fermions and bosons, and with an external constant R-symmetry 
gauge field~$A^{R}$ turned on. In the GLSM description, this background gauge field couples to all the fields 
carrying non-zero R-charges via their covariant derivatives, consistent with the description of the elliptic 
genus as a partition function with twisted boundary conditions \eqref{defell}\footnote{In particular, this definition 
of the path integral includes non-linear (in this case quadratic) self-couplings of the field $A^{R}$, as we shall see below. 
This is related to the fact that the elliptic genus is not modular invariant but only covariant. The elliptic genus, a 
Jacobi form of weight zero and non-zero index, gains a prefactor under modular transformations -- see Eqn.~\eqref{modtransform}. 
A path integral with linear couplings 
of the type~$\chi(\t,z) = \int [D \v] [D \psi] \exp\big(-S[\v,\psi]- \int A^{R\mu} \,J^{R}_{\mu}[\v,\psi] \big)$,
on the other hand, is invariant under modular transformations, and differs from the elliptic genus by a 
prefactor~$e^{\pi m z^{2}/\t_{2}}$, as described in~\cite{Kraus:2006nb}. }.

It is convenient to introduce the complex parameters~$u,z$ that correspond to the holonomies of the dynamical~$U(1)$ 
gauge field~$A$ and the constant external gauge field~$A^{R}$ that couples to the left-moving~$R$-current: 
\be \label{defuz}
u =\oint_A A  - \tau \oint_B A  \, , \qquad z = \oint_A A^\text{R} - \tau \oint_B A^\text{R} \, ,
\ee
where~$A$ and~$B$ refer to the time and space circles of the torus respectively. 
Since shifting the gauge field by an integer is a symmetry of the functional integral, these parameters 
are defined on the torus~$E(\t) = \IC/(\IZ \t + \IZ)$. 
The parameter~$u$ is manifestly complex, with its two real parameters corresponding to the 
holonomies of the gauge field on the two independent cycles of the torus. The parameter~$z$, on the other 
hand, corresponds to the twist on the spatial circle in the Hamiltonian description, which is a priori real. 
Complex~$z$ is taken to mean to be an analytic continuation of this twist parameter\footnote{We thank Sungjay Lee
for discussions on this matter.}.

The result of~\cite{Benini:2013nda} for the elliptic genus of a compact GLSM is a two-dimensional integral 
over~$(u, \ubar)$ on the torus~$E(\t)$.  As we shall review below, 
the integrand is a total derivative in~$\ubar$ and the integral reduces, via Cauchy's theorem, to a contour 
integral in~$u$ of a meromorphic function of~$u$. 
The answer, which can then be expressed as a sum over residues of this meromorphic function, 
can be easily seen to be a holomorphic function of~$\tau$. 
In our case of non-compact GLSMs, the integrand is not a total derivative in~$\ubar$, and we are 
left with a two dimensional integral over~$(u, \ubar)$ that gives us the non-holomorphic function~\eqref{chicig}.

In subsection~\S\ref{hom}, we review the method of~\cite{Benini:2013nda} for compact theories 
with abelian gauge groups, spelling out the details that are relevant to us. 
In subsection \S\ref{inhom}, we shall adapt the same method to the case of the non-compact 
theories that interest us.

\subsection{A review of the compact case \label{hom}}

The field content in this case is:
\begin{itemize}
\item A vector multiplet as in~\S\ref{GLSMCig}, with supersymmetry variations:
\bea \label{susyvec}
\d \l_{+} & = & i \ve_{+} \, D + \p_{+} \bsigma \, \ve_{-} - F_{01} \, \ve_{+} \, , \\ 
\d \l_{-} & = & i \ve_{-} \, D + \p_{-} \sigma \, \ve_{+} + F_{01} \, \ve_{-} \, ,  
\eea
and their complex conjugate equations. 
The action of the vector multiplet, invariant under the above supersymmetry transformations is:
\be \label{vecact}
S_{\rm vec} = {1\over 4\pi}\int\dd^2x~ {1\over 2e^2}\Bigl(
-\partial^{\mu}\bsigma\partial_{\mu}\sigma
+i\blambda_-(\partial_0+\partial_1)\lambda_-
+i\blambda_+(\partial_0-\partial_1)\lambda_+
+F_{01}^2+D^2\Bigr) \, .
\ee
\item 
Chiral multiplets~$\Phi_{i}, (i=1, \cdots, N)$ coupled to the gauge multiplet with charge~$Q_{i}$,
and vector $R$-charge~$R_{i}$ for the bottom component~$\phi_{i}$, 
with supersymmetry variations:
\bea \label{susychiral}
\d \psi_{i+} & = & i \, \CD_{+} \phi_{i} \, \overline \ve_{-}  + \sqrt{2} F_{i} \, \ve_{+} 
- 2 Q_{i} \, \phi_{i} \, \bsigma \, \overline \ve_{+} \, , \\  
\d \psi_{i-} & = & - i \, \CD_{-} \phi_{i} \, \overline \ve_{+}  + \sqrt{2} F_{i} \, \ve_{-} 
+ 2 Q_{i} \, \phi_{i} \, \bsigma \, \overline \ve_{-} \, ,
\eea
and their complex conjugate equations. 
The action of the chiral multiplet coupled to the vector multiplet, invariant under the above supersymmetry transformations, is 
\bea \label{chiract}
S_{\Phi_{i}} = &&\!\!\!\!{1\over 4\pi}\int\dd^2x~\Bigl(
-\CD^{\mu}\bphi_{i}\CD_{\mu}\phi_{i}  +i\bpsi_{i-}(\CD_0+\CD_1)\psi_{i-}
+i\bpsi_{i+}(\CD_0-\CD_1)\psi_{i+} \nonumber  \\
&&~~~~~~~~~~~~~~~
+D|\phi_{i}|^2+|F_{i}|^2 -|\sigma|^2|\phi_{i}|^2
-\bpsi_{i-} \sigma\psi_{i+} -\bpsi_{i+} \bsigma\psi_{i-} \nonumber  \\
&&~~~~~~~~~~~~~~~
-i\bphi\lambda_-\psi_{i+} 
+i\bphi\lambda_+\psi_{i-}  +i\bpsi_{i+} \blambda_-\phi_{i} -i\bpsi_{i-} \blambda_+\phi_{i} \Bigr) \, . 
\eea
The covariant derivatives are $\CD_{\mu}\phi_{i} = (\p_{\mu} + i Q_{i} v_{\mu})\phi$ and 
$\CD_{\mu}\psi_{\pm} = (\p_{\mu} + i Q_{i} v_{\mu})\psi_{\pm} $.  
\end{itemize}

The action of the theory of a $U(1)$ vector multiplet coupled to~$N$ chiral multiplets is:
\be
S = S_{vec} + \sum_{i=1}^{N} S_{\Phi_i}
\ee 
In addition to the coupling given in~\eqref{vecact},~\eqref{chiract}, all the fields couple to the external $R$-symmetry gauge 
field via their R-charges.
The R-charges of the bottom components of the superfields are~$R_{i}$ for the chiral multiplets and~$0$ for the vector multiplet. 
The charges of the other fields of the multiplets are determined by the the superalgebra, with the left-moving 
supercharges~$(\CQ_{-}, {\overline \CQ_{-}})$ having charges~$(-1,+1)$, and the right-moving supercharges having 
vanishing charges. 

To apply the localization technique, we follow the general treatment of~\cite{Pestun:2007rz}. 
We first need to choose a supercharge that is a symmetry of the action and that annihilates the 
observable that we want to compute. For this we choose the right-moving 
supercharge~$\CQ = \CQ_{+}+\overline\CQ_{+}$ that obeys the algebra
\be
\CQ^{2} = {\overline L_{0}} \, . 
\ee

The first step in supersymmetric localization is to deform the action by a~$\CQ$ exact term of the 
form~$\lambda \CQ V$, where~$V$ is a fermionic operator invariant under~${\overline L_{0}}$. 
A nice choice is to pick
\be \label{defV}
V=\sum_{\psi=\psi_{i\pm}, \l_{\pm}} \int \dd^{2}x \, \bpsi (x) \CQ \, \psi (x) \, .
\ee 
On taking~$\lambda \to \infty$, the functional integral localizes to the~$\CQ$-invariant configurations of the theory.  
The result is an integral over the $\CQ$-invariant configurations of the action of the theory (the localization locus) 
times a one-loop determinant coming from the fluctuations in the directions normal to the localization locus:

\subsubsection{The localization locus \label{locloc}}

With the above choice of~$\CQ$, we need to put~$\ve^{+}=1$ and $\overline \ve^{+} =1$ in the supersymmetry 
variations~\eqref{susyvec} and~\eqref{susychiral} and set them to zero. These equations imply:
\bea \label{loceqns}
 \p_{+} \s = \p_{+} \bsigma =0 \, ,&& \qquad D = F_{01} =0 \, , \\
D_{+} \phi_{i} = D_{+} \bphi_{i} =0 \, ,&& \qquad F_{i} = \bF_{i} = 0 \, , \\ 
\phi_{i} \, \bsigma && = \bphi_{i} \, \s=0  \, .
\eea
In the Euclidean theory, the equations~$\p_{+} \s = \p_{+} \bsigma =0$ imply that~$\sigma$ is a constant. 
Similarly, the equations~$D_{+} \phi_{i} = D_{+} \bphi_{i} =0$ on a torus with non-zero gauge field~$A$ 
imply that~$\phi=0$. 
This automatically solves the third line of equations. 
The full set of solutions is therefore parameterized by the complex constant parameterizing the zero 
mode~$\s=\sigma_{0}$, and the zero modes of the gauge field. 
On a torus the gauge field zero modes are parameterised by the Wilson lines along the two directions of the 
torus, labelled by the complex parameter~$u$ defined in~\eqref{defuz}, that takes values in the 
torus~$E(\t)=\IC/(\IZ \tau + \IZ)$. Note that the measure of integration is~$\frac{\dd^{2}u}{\t_{2}}$ 
which is invariant under the translations by~$\IZ \tau + \IZ$, so the integral over~$E(\t)$ of a function invariant 
under shifts of the lattice~$\IZ \tau + \IZ$ is well-defined.

An equivalent way to reach the above conclusions is to notice that the actions~\eqref{vecact} and~\eqref{chiract} are 
actually~$\CQ$-exact quantities~\cite{Benini:2013nda} and so we can work at any value of the coupling constants of the action.
Thus we recover, in the path-integral formalism, the Hamiltonian statement that the elliptic genus is independent 
of the parameters of the action. 
When~$e \to 0$, we obtain the elliptic genus of the UV GLSM, while when~$e \to \infty$, we should obtain the 
elliptic genus of the IR interacting SCFT. Working at~$e \to 0$, we can simply minimise the free Euclidean action 
of the vector multiplet. Similarly, we can analyze the free action of the chiral multiplet. This gives the same solutions 
as above, namely the locus is parameterized by the two complex zero modes~$(u,\s)$.

In fact, the locus can be reduced further for the specific problem of computing the elliptic genus. Note that the 
path integral has an additional term which is the coupling of the right-moving R-charge to the chemical potential~$z$.
This chemical potential can be thought of as a background value of a gauge field. 
As shown in Equation~\eqref{Qexcur}, the right-moving part of the R-current of the vector multiplet 
is also~$\CQ$-exact, and one can add this to the free action of the vector multiplet. Note that both these 
terms scale as~$1/e^{2}$, and so it is 
consistent to keep these two terms as~$e \to 0$. The additional term has the effect of adding a mass 
proportional to~$z$ to~$\s$, and therefore lifts the zero mode of~$\s$. 
The same term also gives a mass to the right-moving fermion~$\l_{+}$ proportional to~$z$. 
These facts will be useful in our further analysis. 

The bottom line of the above analysis is that the localization locus is paramterized by the complex 
parameter~$u \in E(\t)=\IC/(\IZ \tau + \IZ)$ corresponding to the Wilson lines of the gauge field around the torus, 
with all other fields vanishing. 

\subsubsection{The fluctuation integral}

The next step in the localization procedure is to compute the one-loop determinant of the quadratic fluctuations 
of the fields that are orthogonal to the localization locus. In this case, this means we have to compute the 
integral over quadratic fluctuations of all the fields in the theory with a background value of the gauge field 
zero mode~$u$. In doing this integral, one has to be careful with the treatment of various zero-modes of 
fields with kinetic terms\footnote{The fields that do not have kinetic terms can be integrated exactly, and will not 
cause problems.}, for a naive treatment 
bosonic or fermonic zero-modes will lead to divergences or zeros that dramatically affect the answer. 

In the vector multiplet, we have the 
gauge field zero mode~$u$ that we keep explicit in the analysis, and  the zero modes of the left-moving 
gaugini~$(\l_{-},\blambda_{-})$. 
The other fields~$(\s,\l_{+})$ and their complex conjugates are charged with respect to the R-symmetry gauge field
and thus do not have zero modes. 
In the chiral multiplet, all the dynamical fields are charged with respect to the~$U(1)$ gauge field, so it seems that 
none of them have a zero mode. However, there is a subtlety in the last statement 
because the boson~$\v_{i}$ can become massless when the condition 
\be
\label{sing}
Q_i u + \frac{R_i}2 \,z  = 0 \pmod{\IZ\tau + \IZ}  
\ee
is fulfilled. This can happen at a finite number of points on the torus~$E(\t)$, the set of which is called~$M_{\rm sing}$. 
The condensation of this boson can cause a potential divergence in the path-integral. The authors 
of~\cite{Benini:2013nda} argue that any such divergence is eliminated from the integral in a natural 
way if one follows their prescription. 

According to this prescription, one should first perform the~$u$ integral at a small but finite value of coupling 
constant~$e$ -- this implies keeping the integral over the constant mode of the auxiliary field~$D$ in the vector 
multiplet. The integral is then defined by cutting out a small disk of size~$\ve$ around each potentially singular 
point, and the authors of~\cite{Benini:2013nda} show that this integral has a finite limit 
as~$\ve\to 0$\footnote{See Appendix A of~\cite{Harvey:2013mda} for a similar prescription in a closely 
related context.}. One can then perform the~$D$-integral and finally take the limit~$e \to 0$. 

Denoting this sequence of limits by $\lim_{\e,\varepsilon\to0}$, the formula for the elliptic genus is:
\be \label{intchilim}
\chi(\t,z) = \lim_{\e,\varepsilon\to0} \int_{\IR} \dd D
\int_{E^{\ve}}  \frac{\dd^2u}{\t_{2}}  \, \exp\Big( - \frac1{2 e^2}D^2  \Big)  \, f_e(u,D) \, \;.
\ee
Here~$E^{\ve}$ is the torus defined by the modular parameter~$\t$ with disks of size~$\ve$ excised around 
each point ~$u \in M_{\rm sing}$ defined after~\eqref{sing}. The function~$f_e(u,D)$ is the result of the 
path integral over all fields in the weak coupling limit, as a function of~$u$ and~$D$. 
For models considered in~\cite{Benini:2013nda}, the no-chiral-anomaly condition~$\sum_{i} Q_{i} =0$ 
implies that the function~$f_{e}(u)$ above is invariant under the elliptic transformations~$u \to \l \t + \mu$, $\l, \mu \in \IZ$.
This is a consistency condition that allows us to integrate such a function on the torus~$E(\t)$.

Thus we need to compute the one-loop fluctuation determinants of the quadratic operators acting on the 
various bosons and fermions in the theory with~$z$,~$u$ and~$D$ turned on. The solution to this problem 
is well-known (see e.g.~\cite{Ginsparg:1988ui,AlvarezGaume:1986es}) and involves the evaluation of infinite 
products of the form
\be
\prod_{m,n} (m+n\tau+u)
\ee
for left-moving fermions, its complex conjugate for right-moving fermions, and
\be \label{bosdet}
\prod_{m,n} \frac{1}{|m+n\tau+u|^2 + i D}
\ee
for bosons. 
Here we continue to follow the conventions of~\cite{Benini:2013nda}, which involves a Euclidean rotation 
and a rescaling of~$D$ by~$\pi/\tau_2$ compared to the usual conventions, say of~\cite{Witten:1993yc}.

The contribution from a chiral multiplet $\Phi$ of vector-like R-charge~$R$ and gauge charge~$Q$ is 
\be
Z_{\Phi,Q}(\tau,z,u,D) = \prod_{m,n} \frac{\big( m+n\tau+(1- \tfrac R2 ) z - Q u \big) 
\big( m + n \bar\tau + \frac R2 \bar z + Q \bar u \big)}{ \big| m + n\tau + \frac R2 z + Qu \big|^2 + i QD} \, . 
\ee
When $D=0$, this simplifies to
\be
\label{chiraldet}
Z_{\Phi}(\tau,z,u)=\frac{\vth_1(\t,(R/2-1)z + Qu) }{\vth_1(\t,Rz/2 + Qu)} \;.
\ee
The function $\vth_1(\t,z)$ is the Jacobi theta-function that is odd in~$z$. 
The contribution from a~$U(1)$ vector multiplet  is:
\be
\label{vectordet}
 Z_{\rm vec}(\tau,z)= -\frac{\eta(\t)^3}{\vth_1(\t,z)} \, . 
\ee
We have summarized the definitions and conventions for the Jacobi theta and Dedekind eta function in 
Appendix~\S\ref{JacApp}. One property that plays a role in the following is that the Jacobi theta 
function~$\vth_{1}(t,z)$ is an odd function of~$z$ and has a simple zero at~$z=0$. 

The one-loop determinants \eqref{chiraldet}, \eqref{vectordet} summarize the participating non-zero modes. 
In the vector multiplet, the dynamical fields are a complex scalar $\s$ (left- and right-moving) 
and a left- and right-moving complex fermion~$\l_{\pm}$. The boson does not have any zero 
mode, and neither does the right-moving fermion. The left-moving fermion has a complex zero mode that we 
took out of this computation, we shall discuss this separately shortly. Correspondingly, in the fermonic 
determinant (the numerator of~\eqref{vectordet}), we need to divide out by the simple zero at~$z=0$ of the 
function~$\vth_{1}(\t,z)$~\eqref{etatheta}. In the chiral multiplet, the dynamical fields are a complex 
boson~$\phi$ and a complex fermion~$\psi_{\pm}$. All the oscillator modes are present, but we can see 
that for~$u\in M_{\rm sing}$, $Z_{\rm vec}$ diverges.

The final ingredient is the integration over the zero mode of the complex gaugino~$\l_{-}$ which needs 
to be saturated. Since~$\l_{-}^{0}$ is a complex zero mode of the kinetic operator in~\eqref{vecact}, its 
action is independent of~$e$, and its only contribution comes from its coupling to the chiral multiplet 
in~\eqref{chiract}. The effect of the integration is to pull down one factor of the fields that it couples to. 
This gives rise to 
\be \label{lambdazm}
f_e(u,D) = \bigg\langle \int \dd^2x\, \sum_i Q_i \bar\psi_i \phi_i  \int \dd^2x\,  \sum_i Q_i \psi_i \bar\phi_i \bigg\rangle \, . 
\ee

For the computation below, we will need to evaluate this in the~$e \to 0$ limit, i.e.~one can simply 
evaluate~\eqref{lambdazm} in the free theory.  For $u \in E^{\ve}$ and for $D$ whose imaginary part is 
close enough to $0$, we have  
\be
\label{hg}
f_e(u,D) \xrightarrow[e\,\to\,0]{} h(\tau,z,u,D) \, g(\tau,z,u,D) \,,
\ee
where 
\be \label{defg}
g(\tau,z,u,D) = Z_{\rm vec}(\t,z)  \prod_i Z_{\Phi,Q_i}(\tau,z,u,D) 
\ee
is the one-loop determinant without the zero modes, and 
\be \label{defh}
h(\tau,z,u,D)=
- \frac{i}{\pi} \sum_{i,n,m} \frac{Q_i^2}{\displaystyle
\left( \left| m + n \tau + Q_i u + \tfrac{R_i}2 z \right|^2+i Q_iD \right)
\left( m + n \bar\tau + Q_i \bar u + \tfrac{R_i}2 \bar z \right)} \, . 
\ee
One can check that 
\be \label{heqdg}
h(\tau,z,u,D) \, g(\tau,z,u,D)=  - \frac1{\pi D} \,
\frac{\partial}{\partial\overline{u}} g(\tau,z,u,D) \, .
\ee
From the explicit expressions for the determinants~\eqref{chiraldet}, \eqref{vectordet}, we can check, using the 
Jacobi properties, that the condition~$\sum_{i} Q_{i}=0$ implies that the function~$g(u)$ is invariant under 
shifts~$u \to u + \l \t + \mu$, $\l, \mu \in \IZ$, and so the integral over the torus~$E(\t) = \IC/(\IZ \t + \IZ)$ is 
well-defined, i.e.~independent of the coset representative.

\subsubsection{Reduction to a contour integral}

Putting in all these ingredients into the integral~\eqref{intchilim}, one gets an integral: 
\be \label{chiint1}
\chi(\t,z) = \lim_{\e,\ve\to0}
\int_{\Gamma_-}\dd D\, \t_{2} \exp \Big( - \frac1{2e^2}D^2 \Big)
\int_{E^{\ve}}  \frac{\dd^2u}{\t_{2}} \, h(\t,z,u,D) \, g(\t,z,u,D) \;.
\ee
The factor of~$\t_{2}$ in the measure arises because of the rescaling mentioned below~\eqref{bosdet}. 
We have also pulled the~$D$-integral away from the real axis to a contour that runs just below the 
real axis close enough to it so as to avoid all the poles i.e.~$D \in \IR \pm i\delta$ with $0<\delta\ll \varepsilon^2$.
(The final answer will be independent of such a~$\d$.)

Using~\eqref{heqdg}, \eqref{chiint1}, and Cauchy's theorem, we get:
\be
\chi(\t,z) = - \lim_{\e,\ve\to0}
\int_{\Gamma_+}\dd D\, \frac1{2\pi i D} \exp \Big( - \frac1{2e^2}D^2 \Big)
\oint_{\p E^{\ve}}\dd u\,  g(\t,z,u,D) \;.
\ee
We have thus expressed the elliptic genus as the result of two contour integrals~$D$ and~$u$. The final step in 
the manipulations is a deformation of the~$D$ contour that we shall not spell out in detail (see Section~3 
of~\cite{Benini:2013nda}). Following those steps, the~$D$-integral simply picks up the residue at~$D=0$, and 
the~$u$-integral becomes a sum over closed contours encircling~$\{u \in M_{\rm sing}/ R_{i}>0\} \equiv \{ u \in M_{\rm sing}^{+} \}$:
\be
\chi(\t,z)  = - \oint_{u \in M_{\rm sing}^{+}} {\dd u}\,  g(\t,z,u,0) \;.
\ee

Note that this slightly lengthy procedure above of treating the various zero modes separately is necessary, 
otherwise we end up with the wrong two-dimensional integral over~$u$ instead of the correct contour integral in~$u$.

\subsection{The inhomogeneous non-compact theories \label{inhom}}

Now we would like to consider a theory which has, in addition to a~$U(1)$ vector multiplet and chiral 
multiplets~$\Phi_{i}$ as in the previous section, a chiral compensator multiplet~$P$, as in~\S\ref{GLSMCig}, which 
transforms inhomogeneously under super gauge transformations~$\Lambda$ as 
\be \label{ggetransP}
P \to P + i \Lambda \, .
\ee

The action of the~$P$ multiplet coupled to the gauge multiplet is: 
\bea \label{actphi}
S_{P} && = {1\over 4\pi}\int\dd^2x~{k\over 2}\bigl(-\CD^{\mu}\bp\CD_{\mu}p +i\bchi_-(\partial_0+\partial_1)\chi_-
+i\bchi_+(\partial_0-\partial_1)\chi_+ +D(p+\bp)+|F_P|^2 \nonumber \\
&&~~~~~~~~~~~~~~~
-|\sigma|^2 +i\chi_+\lambda_-  -i\chi_-\lambda_+
+i\bchi_+\blambda_- -i\bchi_-\blambda_+\bigr) \, . 
\eea
Recall that $\CD_{\mu}p =\partial_{\mu}p+iv_{\mu}$, and that the imaginary part of~$P$ is periodically 
identified with periodicity~$2 \pi$.
The full action of the theory is:
\be \label{fullact}
S = S_{vec} + \sum_{i=1}^{N} S_{\Phi_i} + S_{P} \, .
\ee 
As we saw in~\S\ref{GLSMCig}, the conserved right-moving~$R$-symmetry in the quantum theory is 
\be \label{RcurP}
\wt j_R^+= j_{R,\Phi}^{+} + \frac{k}2 \, \chi_-\bchi_- - 2\p_{-} \Im \, p - 2 v_{-}\, .
\ee

Now we would like to compute the path integral with the inclusion of the~$P$-multiplet. 
We follow the same route as in the previous section using localization with respect to the same 
supercharge~$\CQ$, but there are crucial differences in the details of the two computations. 
The first difference is that the action of the~$P$-multiplet is not~$\CQ$-exact due to the presence of 
total-derivative terms as mentioned in~\cite{Hori:2001ax, Ashok:2013zka}, so we cannot naively minimize 
the action of the~$P$-multiplet.

In order to proceed, we observe that since the vector multiplet action is~$\CQ$-exact, we have already 
reduced the problem to an evaluation of the path-integral at~$e \to 0$ in the sense discussed in the last section. 
If we do the vector multiplet integral first, this means that we are left with an essentially \emph{free}~$P$-multiplet. 
The only couplings we need to keep while computing the integral over the~$P$-multiplet fields are the coupling to 
the zero modes gauge fields (that do not have a $1/e^{2}$ self-coupling) i.e.~the complex zero mode of the 
gauge field~$u$ and the complex zero mode of the left moving gaugino~$\l_{-}$. The action of these fields is:
\bea \label{actphiquad}
S^{\rm free}_{P} && = {1\over 4\pi}\int\dd^2x~{k\over 2}\bigl(-(\p^{\mu}\bp-iu^{\mu}) 
(\p_{\mu}p + i u_{\mu}) +i\bchi_-(\partial_0+\partial_1)\chi_-
+i\bchi_+(\partial_0-\partial_1)\chi_+  \nonumber \\
&&~~~~~~~~~~~~~~~
+D(p+\bp) +|F_P|^2 +i\chi_+^{0}\lambda_-^{0}  +i\bchi_+^{0}\blambda_-^{0} \bigr) \, . 
\eea
Note that there are zero modes for the right-moving fermion~$(\chi_{+},\bchi_{+})$, but not for the 
left-moving fermion~$(\chi_{-},\bchi_{-})$ because of its coupling to the R-symmetry gauge field 
via the current~\eqref{RcurP}.

The expression for the elliptic genus is as in~\eqref{intchilim}: 
\be \label{chiforP}
\chi(\t,z) = \lim_{\e,\varepsilon\to0} \int_{\G^{-}} \dd D \, \t_{2}
\int_{E^{\ve}}  \frac{\dd^2u}{\t_{2}}  \, \exp\Big( - \frac1{2 e^2}D^2 \Big)  \, f^{(P)}_e(u,D) \, ,
\ee
where the function~$f^{(P)}_e(u,D)$ is taken to mean the integral over all the fields except~$u$
and~$D$. We include a superscript to indicate that the integration over the~$P$-multiplet gives a different 
function than the one in the previous section. 

As before, we first perform the integral over the fermion zero modes, but the answer 
is now different because of the existence of new fermion zero modes~$(\chi_{+},\bchi_{+})$, and the coupling 
of these modes to~$(\l_{-},\blambda_{-})$ in~\eqref{actphiquad}.  Integrating over all these fermion zero modes, we 
simply get a factor of one in the integral. We are then left with the one-loop determinants of all the non-zero modes 
in the theory as a function of~$u$ and~$D$. We have already computed the one-loop determinants for the 
fields of the chiral multiplets~$\Phi_{i}$ and the gauge multiplet. It remains to do so for the fields of the~$P$-multiplet.

In the~$P$-multiplet, the left-moving fermion~$(\chi_{-},\bchi_{-})$ 
is charged under the $R$-symmetry gauge field and it is not coupled to the~$U(1)$ gauge field nor to~$D$. 
The non-zero modes of the boson do not couple to either of the gauge fields nor to~$D$, 
the zero mode of~$\Re(p)$ couples to~$D$ but not to~$z$ or~$u$, while~$\v_{P}= \sqrt{2k} \, \Im(p)$ 
with no oscillators, couples to~$u$ and~$z$ but not to~$D$.  
This mode has canonical kinetic term and lives on a circle with radius~$R=\sqrt{2k}$, i.e.~for~$m,w \in \IZ$, 
we have the periodic identification:
\bea
\v_{P}(x^{1}+2 \pi,x^{2}) &=& \v_{P}(x^{1},x^{2}) + 2 \pi w \sqrt{2k} \, , \nonumber \\
\v_{P}(x^{1}+2 \pi \t_{1},x^{2}+2 \pi \t_{2}) &=& \v_{P}(x^{1},x^{2}) + 2 \pi m \sqrt{2k} \, .
\eea
This periodicity allows for the following solutions to the free quadratic action~\eqref{actphiquad}:
\be
\v_{P}(x_{1},x_{2}) =x^{1} w \sqrt{2k} + x^{2} (m-w\t_{1})\sqrt{2k}/\t_{2} \, ,
\ee
that corresponds to momentum and winding in the target space. 
Summing over the contributions of these modes, and taking into account the free oscillator modes, we 
obtain the one-loop determinant of the~$P$-multiplet: 
\be
\label{Pdet}
Z_{P}(\tau,z,u)=-k\, \frac{1}{D \t_{2}} \, \frac{\vth_1(\t,z) }{\eta(\t)^{3}} \, 
\sum_{m,w \in \IZ} \exp \Big(- \frac{\pi k}{\t_{2}} \big(m + w \t + u+\frac{z}{k}\big)(m + w \tbar + \ubar + \frac{z}{k}) \Big) \, , 
\ee
where the factor of~$1/D$ comes from integrating over the zero mode of~$\Re(p)$. 
%Here the parameter~$z$ is real, consistent with a real twist in the boundary 
%conditions in the Hamiltonian description, as discussed below~Eqn.~\eqref{defuz}. 
%%In the next section, we shall discuss the analytic continuation 
%%to arbitrary complex parameter~$z$. 
Note that~$\v_{P}$ undergoes a translation under the action of the R-symmetry \eqref{RcurP}, 
which accounts for the fact 
that~$z$ appears in both the left and the right-moving part of the exponential above. 
%explained below Eqn.~\eqref{defuz}.

We have to multiply this expression with the vector multiplet determinant and the chiral multiplet determinant, 
and integrate over the parameter~$u$. However, there is another subtlety here coming from the fact that 
the correct gauge field 
background for the chiral multiplet is now the modified expression~$A_{\mu} = v_{\mu}+\p_{\mu} \Im \, p$. In 
this background we obtain chiral determinant to be:
% Wilson line is gauge invariant now. 
\be \label{chiraldetnew}
Z_{\Phi}(\tau,z,u)=\frac{\vth_1(\t,(R/2-1)z + Q(u+m+w\t))}{\vth_1(\t,Rz/2 + Q(u+m+w\t))} \, .
\ee
Using the Jacobi property~\eqref{elliptic} of the~$\vth$ function, we get:
\be \label{chiraldetnew1}
Z_{\Phi}(\tau,z,u)=e^{2 \pi i Q zw} \frac{\vth_1(\t,(R/2-1)z + Qu)}{\vth_1(\t,Rz/2 + Qu)} \, .
\ee
The above argument is equivalent to explicitly considering the anomalous transformation behavior of 
charged fermions in a gauge field background as mentioned in~\cite{Ashok:2011cy}. The phase factor 
that depends on~$w$ is also consistent with the elliptic symmetry of~$u$,~i.e. the symmetry of large 
gauge transformations of the linear sigma model.

Putting all this into the integral~\eqref{chiforP}, we can perform the~$D$ integral as before to pick up 
the residue at~$D=0$, which gives us:
\bea\label{chifinal}
\chi_{N}(\t,z) \= && k \lim_{\varepsilon\to0}  \int_{E^{\ve}} \frac{\dd^2u}{\t_{2}} \, 
\prod_{i=1}^{N} \frac{\vth_1(\t,(R_{i}/2-1)z + Q_{i}u) }{\vth_1(\t,R_{i}z/2 + Q_{i}u)} \, \times \cr
&& \qquad \qquad \qquad \qquad 
\sum_{m,w \in \IZ} \ e^{2 \pi i (\sum_{i}Q_{i}) z w - \frac{\pi k}{\t_{2}} 
(m + w \t + u+\frac{z}{k})(m + w \tbar + \ubar + \frac{z}{k})} \, .
\eea
Note that the above integrand is also invariant under the elliptic transformations~$u \to u + \l \t + \mu$,
$\l, \mu \in \IZ$, so that the integral is well-defined on the coset~$E=\IC/(\IZ \t + \IZ)$. Demanding this 
invariance is another way to determine the modified chiral determinant~\eqref{chiraldetnew1}.
This fact was already emphasized in~\cite{Ashok:2011cy} for the cigar coset theory. It was also shown in that paper 
that~$\chi(\t,z)$ is modular invariant.  
When all the~$R_{i}=2\frac{1-N}{k}$ and~$Q_{i}=1$, we get the expression in~\cite{Ashok:2013zka}. 
For generic~$N>1$, the expression~\eqref{chifinal} has potential singularities 
that we briefly discuss in \S\ref{holanom}. 
When we have only one field~$\Phi = \Phi_{1}$ with~$R=0, Q=1$, we get: 
\be\label{chifinalcig}
\chi^{\rm cig}(\t,z) \= k \lim_{\varepsilon\to0}  \int_{E^{\ve}} \frac{\dd^2u}{\t_{2}} \, 
 \frac{\vth_1(\t,-z + u) }{\vth_1(\t,u)} \, 
\sum_{m,w \in \IZ} \ e^{2 \pi i z w - \frac{\pi k}{\t_{2}} (m + w \t + u+\frac{z}{k})(m + w \tbar + \ubar + \frac{z}{k})
} \, . 
\ee

The parameter~$z$ in the above discussion is a priori real, consistent with a real twist in the boundary 
conditions in the Hamiltonian description, as discussed below~Eqn.~\eqref{defuz}. 
We can analytically continue this result to complex values of~$z$. 
In the next section we show that the above expression for the elliptic genus obeys the modular, elliptic, and 
holomorphicity properties as expected\footnote{We thank the referee for pointing out an 
error in the above formula in a previous version of this paper.} for an arbitrary complex parameter~$z$.  
For now we note that, by the periodicity of the integrand under~$u \to u+\l \t + \mu$, the expression~\eqref{chifinalcig} 
is equivalent to the expression~\eqref{chicig} for real values of~$z$ as expected. 
The shift of variables~$u\to u+z/k$ to go between the expression~\eqref{chifinalcig} and~\eqref{chicig} is a 
reflection of the fact that the free-fermions of the IR coset theory are not the same as those of the UV theory. 
The fermions of the cigar coset have $R$-charge~$1+1/k$ while those of the UV free theory
have the R-charge~$1$. 
The correct identification of the UV and IR theories goes via the identification of the $R$-symmetry which is 
conserved throughout the RG flow. 

In the expression~\eqref{chifinal}, the factors coming from the non-zero modes of the $P$-multiplet 
cancel those coming from the vector multiplet.
We are left with the holomorphic contribution of the chiral multiplets~$\Phi_{i}$, $(i=1,\cdots,N)$, 
along with the non-holomorphic contribution of the compact compensator boson~$\v_{P}$ through 
its winding and momentum modes. The observation that the non-holomorphicity
of the elliptic genus of the cigar comes from the winding and momenta of a free boson has been made  
in~\cite{Ashok:2011cy,Eguchi:2010cb,Ashok:2012qy,Ashok:2013kk}.
The conceptual addition we make to this is the identification of this boson as the compensator 
(or equivalently, as the St\"uckelberg field) for the anomalous chiral rotations of fermions
in the class of theories studied in this paper. 

One can give a simple interpretation of~\eqref{chifinal} in the IR theory as well. 
If we integrate out the gauge field in a naive way, we are left with the fields~$\Phi_{i}$ that 
are the coordinates of a target space which is not Ricci-flat. The oscillator modes of these coordinates 
make up the holomorphic theta functions in the partition function. The non-trivial curvature induces a chiral anomaly as 
in~\eqref{anomalycig}.  The compensator that is needed to cancel the anomaly makes up the remaining contribution.

\section{Modularity and the holomorphic anomaly\label{holanom}}

In the last section, we have derived the elliptic genus of a class of GLSMs labelled by~$N$. This took the 
form of a two-dimensional integral over $u \in E(\t)=\IZ\t+\IZ$ of a non-holomorphic function of~$u$. 
The theory with~$N=1$ is the cigar coset where the integral has been explicitly calculated in terms of 
Appell-Lerch sums. 

For the higher~$N$ cases, we note that the expressions that we obtained~\eqref{chifinal} are formal, 
and in particular the expression develops a singularity when two of the poles in the~$u$-coordinate coincide. 
One can resolve this by introducing more chemical potentials (fugacities) and forcing 
them to take non-coinciding values as done in~\cite{Ashok:2013zka}. 
Even so, the Fourier expansion for the holomorphic part of these expressions is 
not well-defined and experiences a wall-crossing phenomenon. This is unlike the case of the cigar,
and is more reminiscent of the indexed partition functions of the black holes that have been discussed 
in~\cite{DMZ}. It will be interesting to understand the physics of these singularities, and develop the 
mathematical formalism, perhaps along the lines of the formalism developed in~\cite{DMZ}.

In this section, we first briefly present a proof of the modularity and elliptic properties of 
our expression for the cigar elliptic genus. 
%In doing so we present expressions for the elliptic genus equivalent to~\eqref{chifinalcig}. 
We then compute the holomorphic anomaly in the elliptic genera of all the models considered in the previous section, 
and show that this reduces to the known answer for the case of the cigar. 
We hope that this analysis is also useful to understand the higher~$N$ case. 
 
\subsection{Modular and elliptic properties}

The expression~\eqref{chifinalcig} can be rewritten as follows:
\bea\label{chifinalcigagain}
\chi^{\rm cig}(\t,z) &\=& k \lim_{\varepsilon\to0}  \int_{E^{\ve}} \frac{\dd^2u}{\t_{2}} \, 
 \frac{\vth_1(\t,-z + u) }{\vth_1(\t,u)} \, 
\sum_{m,w \in \IZ} \ e^{2 \pi i z w - \frac{\pi k}{\t_{2}} (m + w \t + u+\frac{z}{k})(m + w \tbar + \ubar + \frac{z}{k})
} \cr
&\=& k \lim_{\varepsilon\to0}  \int_{E^{\ve}} \frac{\dd^2u}{\t_{2}} \, \sum_{m,w \in \IZ} \, 
 \frac{\vth_1(\t,-z + u + m + w \t ) }{\vth_1(\t,u+m + w \t )} \; 
e^{-\frac{\pi k}{\t_{2}} (m + w \t + u+\frac{z}{k})(m + w \tbar + \ubar + \frac{z}{k})} \cr
&\=& k  \int_{\IC} \frac{\dd^2u}{\t_{2}}  \, 
 \frac{\vth_1(\t,-z + u) }{\vth_1(\t,u)} \; 
e^{-\frac{\pi k}{\t_{2}} (u+\frac{z}{k})(\ubar + \frac{z}{k})} \, . 
\eea
Here we have used the elliptic property~\eqref{elliptic} of the~$\vth$ function, as in Equations~\eqref{chiraldetnew}, 
\eqref{chiraldetnew1}, to obtain the second line. 
To obtain the third line, we note that the integrand above is invariant under the shifts~$u \to u+ \IZ \t +\IZ$,
and we can therefore exchange the sum over $(m,w) \in \IZ^{2}$ in the integrand with a sum over 
the different coset representatives, thus effectively unfolding the integration region to the whole complex plane. 

We now perform the transformation~$\t \to -\frac1{\t}, z \to \frac{z}{\t}$ on the last line of~\eqref{chifinalcigagain}, 
and accompany it with a change of 
variable~$u=\frac{u'}{\t}$. Using the modular transformation property of the theta functions,
we obtain:
\be
\frac{\vth_1(\t,-z + u) }{\vth_1(\t,u)} \to \frac{\vth_1(-\frac{1}{\t},\frac{-z + u'}{\t}) }{\vth_1(-\frac{1}{\t},\frac{u'}{\t})} 
\= e^{\frac{\pi i}{\t} (z^{2} -2 u' z)} \frac{\vth_1(\t,-z + u') }{\vth_1(\t,u')} \, .
\ee
The exponential factor transforms as:
\be
e^{-\frac{\pi k}{\t_{2}} (u+\frac{z}{k})(\ubar + \frac{z}{k})} \to e^{\frac{2 \pi i z^{2}}{k\t} + \frac{2 \pi i u' z}{\t}}  \, 
e^{-\frac{\pi k}{\t_{2}} (u'+\frac{z}{k})(\ubar^{'} + \frac{z}{k})} \, . 
\ee
The measure of the integral~\eqref{chifinalcigagain} transforms as~$\frac{d^{2}u}{\t_{2}} \to \frac{d^{2}u'}{\t_{2}}$,
and the range of integration of~$u'$  is also the complex plane. 
Putting these facts together, we obtain:
\be
\chi^{\rm cig}(-\frac1{\t}, \frac{z}{\t}) \= e^{\frac{\pi i}{\t} z^{2} (1+\frac{2}{k})} \, \chi^{\rm cig}(\t,z) \, , 
\ee
which is the modular transformation of a Jacobi form of weight 0 and index~$\frac{1}{2}+\frac{1}{k}$, as expected. 

To demonstrate the elliptic property, it is convenient to make a change of variables~$z=kz'$, so that the index 
of~$\v^{\rm cig}(\t,z')\equiv\chi^{\rm cig}(\t,kz')$ is expected to be~$\frac{k^{2}}{2}+k$. To show that this is indeed the case, 
we perform the 
transformation~$z' \to z' + \l \t + \mu$ with~$\l, \mu \in \IZ$, and accompany it with the change of 
variable~$u = u'-\l \t -\mu$ in the last line of~\eqref{chifinalcigagain}. The transformations of the theta functions gives:
\be
\frac{\vth_1(\t,-kz' + u) }{\vth_1(\t,u)} \to e^{-\pi i \l^{2}(k^{2}+2k)\t} \, e^{2 \pi i k \l u' - 2 \pi i k(k+1) \l z'} \, 
  \frac{\vth_1(\t,-kz' + u') }{\vth_1(\t,u')} \, , 
\ee
while the exponential factor transforms as:
\be
e^{-\frac{\pi k}{\t_{2}} (u+\frac{z'}{k})(\ubar + \frac{z'}{k})} \to e^{-2 \pi i k \l u' - 2 \pi i k \l z'}  \, 
e^{-\frac{\pi k}{\t_{2}} (u'+\frac{z'}{k})(\ubar^{'} + \frac{z'}{k})} \, . 
\ee
The form of the measure and the integration region for~$u'$ also do not change. 
Putting these facts together, we obtain:
\be
\v^{\rm cig}(\t, z' + \l \t + \mu) \= e^{-\pi i (k^{2}+2k) \l^{2} \t} \, e^{- 2 \pi i (k^{2}+2k) \l z'}  \, \v^{\rm cig}(\t,z') \, , 
\ee
which is the elliptic transformation of a Jacobi form of index~$\frac{k^{2}}{2}+k$, as expected. 
The analysis of this subsection can be repeated for the elliptic genera of the models~\eqref{chifinal}.

\subsection{The holomorphic anomaly}
 
Starting from the expression~\eqref{chifinal}, we would like to compute~$\p_{\tbar} \chi_{N}(\t,z)$.
Since the measure $d^{2}u/\t_{2}$ is independent of~$\t,\tbar$, we can pull the~$\tbar$-derivative 
inside the integral. We have already seen that the integrand is  invariant under the large gauge 
transformations~$u \to u+\l \t + \mu$, and consequently, it is annihilated by the heat 
operator~$\p_{\tbar} - \frac{i}{4 \pi} \p_{\ubar}^{2}$, as can be verified explicitly. 
Thus we obtain:
\bea
\p_{\tbar} \chi_{N}(\t,z) & = & \frac{ik}{4 \pi}  \lim_{\varepsilon\to0}  \int_{E^{\ve}} \frac{\dd^2u}{\t_{2}} \, 
\p_{\ubar}^{2} \bigg( \prod_{i=1}^{N} \frac{\vth_1(\t,(R_{i}/2-1)z + Q_{i}u) }{\vth_1(\t,R_{i}z/2 + Q_{i}u)} \, 
\sum_{m,w \in \IZ} \, e^{2 \pi i (\sum_{i}Q_{i}) z w - \frac{\pi k}{\t_{2}} |m + w \t + u+\frac{z}{k} |^{2}} \bigg) \nonumber \\
\eea
%A detailed derivation of this method is presented in Appendix A of~\cite{Harvey:2013mda}.

The integrand is a total derivative, so we are left with a boundary integral. The boundary is the union of the 
boundary of the fundamental parallelogram of the~$\IZ \t + \IZ$ lattice and the small circles surrounding the 
singular points~$R_{i}z/2 + Q_{i}u=0$. Since the integrand is periodic in~$u$, the contribution of the first component 
vanishes, and one is left with a contour integral around the points~$u \in M_{\rm sing}$ which can be evaluated
using Cauchy's residue theorem. We obtain:
\bea
\p_{\tbar} \chi_{N}(\t,z) & = & \frac{ik}{4 \pi}  \oint_{M_{\rm sing}}  \dd u \, 
\p_{\ubar} \bigg( \prod_{i=1}^{n} \frac{\vth_1(\t,(R_{i}/2-1)z + Q_{i}u) }{\vth_1(\t,R_{i}z/2 + Q_{i}u)} \, 
\sum_{m,w} \, e^{2 \pi i (\sum_{i}Q_{i}) z w - \frac{\pi k}{\t_{2}} |m + w \t + u+\frac{z}{k} |^{2}} \bigg)  \nonumber  \\
& = & -\frac{k}{4 \pi \t_{2}}  \sum_{i=1}^{n} \frac{\vth_1(\t,z) }{\eta(\t)^{3}}  
\prod_{j \neq i}  
\frac{\vth_1\big(\t,(\frac{R_{j}}{2}-\frac{R_{i}}{2} \frac{Q_{j}}{Q_{i}}-1)z\big) }{\vth_1 \big(\t,(\frac{R_{j}}{2}-\frac{R_{i}}{2}\frac{Q_{j}}{Q_{i}})z\big) } \times \, \nonumber \\
& & \qquad \qquad \qquad \qquad \times \left. \p_{\ubar} \sum_{m,w \in \IZ} \, 
 \, e^{2 \pi i (\sum_{i}Q_{i}) z w - \frac{\pi k}{\t_{2}} |m + w \t + u + z(\frac{1}{k} -\frac{R_{i}}{2Q_{i}}) |^{2}}  
 \right|_{u=-\frac{R_{i}z}{2Q_{i}}} \,\, . 
 \label{freeboson} 
\eea
Here we have assumed that the points of~$M_{\rm sing}$ are all distinct.

For the case of the cigar, we have only one field~$\Phi$ with~$R=0$, $Q=1$, so we get:
\be \label{cigfreebos}
\p_{\tbar} \chi^{\rm cig}(\t,z)  =  -\frac{k}{4 \pi\t_{2}} \, \frac{\vth_1(\t,z) }{\eta(\t)^{3}} \, 
\left. \p_{\ubar} \sum_{m,w \in \IZ} \,  e^{2 \pi i z w - \frac{\pi k}{\t_{2}} |m + w \t + u +\frac{z}{k} |^{2}} \right|_{u=0} \, . 
\ee
After a Poisson resummation of~\eqref{cigfreebos}, we obtain:
\be \label{L0barexpect}
\p_{\tbar} \chi^{\rm cig}(\t,z) =  \frac{i\sqrt{k}}{2\sqrt{\t_{2}}} \, \frac{\vth_1(\t,z) }{\eta(\t)^{3}} \, 
\sum_{n,w \in \IZ} \, (n-wk) \, q^{\frac{(n+wk)^{2}}{4k}} \, \overline{q}^{\frac{(n-wk)^{2}}{4k}}  \,  \zeta^{-\frac{n}{k}+w}  \, . 
\ee
We identity the holomorphic prefactor consisting of a quotient of a theta and eta functions as the partition function
of the field~$\Phi$. The sum over~$(n,w)$ can now be interpreted as a sum over the momenta and winding 
of the compact boson~$\v_{P}$ with momenta and winding around a circle of radius~$R=\sqrt{2k}$.

With a little more work, one can also 
re-express the above holomorphic anomaly equation in the form as written in~\cite{DMZ} 
using the standard~$\vth_{m,\ell}$ functions:
\be \label{holanomchi}
-\frac{2i}{\sqrt{k}} \t_{2}^{1/2} \p_{\tbar} \chi^{\rm cig}(\t,z)  = \frac{1}{k} \frac{\vth_1(\t,z) }{\eta(\t)^{3}} \, 
 \sum_{\a,\b\in \IZ/2k\IZ}  \, e^{2\pi i \frac{\a \b}{k}} \, q^{\frac{\a^{2}}{k}} \, \zeta^{\frac{2\a}{k}} \, 
 \sum_{\ell \mypmod{2k}}  \, \overline{\vth_{k,\ell}^{(1)} (\t)} \; \vth_{k,\ell}\big(\t,\frac{z+\a\t+\b}{k}\big) \, .
\ee
Using~\eqref{holanomA1}, we see that the right-hand side of the above equation is the shadow of the 
Appell-Lerch sum appearing in the cigar elliptic genus~\eqref{ALcig}. 
It would be nice to work out all the details of a similar interpretation for the case~\eqref{freeboson} of many fields.

\section{Discussion \label{discuss}}

The fact that the holomorphic anomaly in the elliptic genus of non-compact theories can be localized to 
the contribution of the compensator multiplet for the chiral anomaly gives us a simple method to pinpoint 
if a given theory has a holomorphic anomaly or not. 
This is most-easily illustrated by considering the mirror dual of the supersymmetric cigar theory, 
i.e.~the~$\CN=2$ supersymmetric Liouville theory~\cite{Hori:2001ax}, which has a condensate of winding modes.
This interacting SCFT has the same field content as the asymptotic cigar theory, i.e.~a complex 
boson~$\rho + i\theta$ and its superpartners~$\psi_{\rho}, \psi_{\theta}$ along with their left-moving 
counterparts. The theory is defined by a superpotential of the type:
\be
\CL^{SL}_{int} = \psi \wt \psi \, e^{-{1 \over Q} (\rho +\wt \rho + i (\theta - \wt \theta ))} + {\rm c.c} \,  ,
\ee
where $\psi = \psi_\rho + i \psi_\theta$ is the superpartner of $\rho + i \theta$ and $\wt \psi$ is its rightmoving 
counterpart. From this superpotential, one can immediately read off the various special features -- firstly, that 
a naive rotation of the fermions is not a symmetry; secondly, when the chiral rotation of fermions is accompanied 
by a shift of the chiral boson, it \emph{is} a conserved symmetry; and finally, that momentum is conserved, 
but winding is not. It is also clear that these three features are related to each other. 

In the mirror cigar picture, the lack of winding is clear geometrically (a winding mode can slide off the tip),
while the anomaly of the chiral rotation is a one-loop quantum effect due to the non-zero Ricci tensor of the cigar. 
Based on these observations, we can say that if the target space for an interacting SCFT has a non-compact 
direction accompanied by a~$U(1)$ isometry 
along which winding is not conserved, the elliptic genus of the theory will suffer from a holomorphic anomaly,
and will have mock modular behaviour.

There are many other GLSMs that flow to non-compact theories that can be studied with our methods,
and are interesting for diverse reasons. 
Among them are orbifolds of the theories studied here and squashed toric models that correspond to 
massive theories like the ``supersymmetric sausage''~\cite{Fendley:1992dm, Aganagic:2001uw}. 
The cigar-like models also have conjectured relations to matrix models~\cite{McGreevy:2003dn, Eguchi:2003ik}. 
Other models correspond to the worldsheet theory of strings in the background of 
NS5-branes in string theory wrapped on various surfaces~\cite{Hori:2002cd}. These intrinsically interesting from 
the string theory point of view, and have found recent applications in the moonshine program~\cite{Harvey:2013mda}. 
Another interesting direction would be to extend our analysis to non-abelian theories as has been 
done for compact models~\cite{Benini:2013xpa}. 

Appell-Lerch sums and related mock modular forms make another interesting appearance in physics 
in the context of the wall-crossing phenomenon for supersymmetric black holes in~$\CN=4$ string theory 
in four dimensions~\cite{DMZ}. In that situation, the result for the supersymmetric index is known, but it is not known 
if the SCFT describing the moduli space of black holes/strings arises as the IR fixed point of any gauged linear
sigma model. It would be very interesting if we can identify these putative UV theories. As remarked in~\S\ref{holanom},
it would be interesting if any of the models with higher~$N$ discussed in this paper is related to 
black hole partition functions.

We believe that our analysis makes some progress to answer the three questions raised in the introduction. 
Finally, a very interesting fourth question is to understand the geometric interpretation of the non-holomorphic 
elliptic genera that we find\footnote{Indeed, this was partly the origin of this investigation. We thank J.~Harvey, 
S.~Katz, and A.~Klemm for discussions on this topic.}. In the compact case, when one considers a~$(2,2)$ 
superconformal field theory with a target space that is in the moduli space of a compact Calabi-Yau manifold, the 
function~$\chi(\t,z)$ coincides with the geometric definition of the elliptic genus of the manifold 
(see~e.g. Equation~(1.1) of~\cite{Gritsenko:1999fk}). 
It would be very interesting if we can find a geometric formula that extends this to include the case of the 
non-compact models discussed here. In this regard, we note that the cigar SCFT is supposed to be the stringy 
description of the region near the singularity of singular Calabi-Yau manifolds in a double scaling 
limit~\cite{Ooguri:1995wj, Giveon:1999zm, Giveon:1999px}. 
Gauged linear sigma models may provide the missing link between non-rational conformal 
field theories and the geometry and topology of non-compact manifolds.

\acknowledgments

We thank Sujay Ashok, Benjamin Assel, Davide Cassani, Atish Dabholkar, Jeff Harvey, Sheldon Katz, 
Albrecht Klemm, Sungjay Lee, Jan Troost, and Bernard de Wit for interesting discussions and comments on issues 
related to those discussed in this paper. We thank the American Institute of Mathematics for the organization  
of the workshop ``Gromov-Witten invariants and number theory'' in April 2013, in which some issues 
discussed in this paper were brought up.

\appendix
\section{Jacobi forms and Appell-Lerch sums \label{JacApp}}

In this appendix, we briefly review the basic facts about Jacobi forms and Appell-Lerch sums that we 
used in the main text. The elliptic genus of a compact Calabi-Yau manifold of complex 
dimension~$d$ is a Jacobi form of weight 0 and index~$d/2$. 
For superconformal field theories, we should replace~$d$
by~$c/3$, and so the index is~$c/6$, where~$c$ is the central charge of the SCFT.

A Jacobi form of weight~$k$ and index~$m$ is a holomorphic function~$\v(\tau, u)$ 
from~$\mathbb{H} \times\IC$ to $\IC$ which is ``modular in $\tau$ and elliptic in $u $'' 
in the sense that it transforms under the modular group as
  \be\label{modtransform}  \v\Bigl(\frac{a\t+b}{c\t+d},\frac{u}{c\t+d}\Bigr) \= 
   (c\t+d)^k\,e^{\frac{2\pi imc u^2}{c\t+d}}\,\v(\t,u)  \qquad \forall \quad
   \Bigl(\begin{array}{cc} a&b\\ c&d \end{array} \Bigr) \in SL(2; \Z) \ee
and under the translations of $u$ by $\mathbb{Z} \tau + \mathbb{Z}$ as
  \be\label{elliptic}  \v(\t, u+\lambda\tau+\mu)\= e^{-2\pi i m(\lambda^2 \t + 2 \lambda u)} \v(\t, u)
  \qquad \forall \quad \lambda,\,\mu \in \Z \, . \ee
We consider $k,m \in \half \IZ$. In addition there are some growth properties that we have 
not mentioned here. We refer the reader to~\cite{Eichler:1985ja} for a nice exposition of the theory of Jacobi forms. 

There is an unfortunate clash of notation between the conventions 
of the Jacobi form literature~\cite{Eichler:1985ja} where~$k$ is used for the weight of the Jacobi form, and in 
the literature on the super Kazama-Suzuki cosets like~$SL(2,\IR)_{k}/U(1)$, for which~$k$ is related to the 
index of the elliptic genus. In the text we have followed the latter convention.

The transformation laws~\eqref{modtransform}, \eqref{elliptic} include the periodicities $\v(\t+1,z) = \v(\t,z)$ and $\v(\t,z+1) = \v(\t,z)$, so $\v$ has a Fourier expansion
\be\label{fourierjacobi} \v(\t,z) \= \sum_{n, r} c(n, r)\,q^n\,\zeta^r\,, \qquad\qquad
   (q :=e^{2\pi i \t}, \; \zeta := e^{2 \pi i z}) \ . 
 \ee

Some interesting functions that appear repeatedly in the main text are the Dedekind eta function, a modular \
form of weight~$1/2$:
\be\label{defeta}
  \eta(\tau)\;:=\; q^{1/24} \prod_{n=1}^{\infty} (1-q^{n}) \, , 
\ee
and the odd Jacobi theta function which is a Jacobi form of weight~$1/2$ and index~$1/2$:
\be
\vartheta_1(\tau,z) = -iq^{1/8} \zeta^{1/2} \prod_{n=1}^\infty (1-q^n) (1-\zeta q^n) (1-\zeta^{-1} q^{n-1}) 
= i \sum_{m \in \ZZ} e^{\pi i (m+\half)}  \, q^{(m+1/2)^2/2} \, \zeta^{m+\half} \, . 
\ee
We have the relation:
\be\label{etatheta} 
 \left. \frac{1}{2\pi i} \, \frac{d}{dz} \vth_{1}(\t,z) \right|_{z=0} \= -i \, \eta(\t)^{3} \, . 
\ee

The Appell-Lerch sum that appears in the main text in \S\ref{ellgencig} is:
\be\label{AP1AP2}  
\AP{1,m}(\t,z) \=  -\frac12 \, \sum_{s \in \mathbb{Z}} q^{ms^2\,}\zeta^{2ms} \, \frac{1 +q^s \zeta}{1 -q^s \zeta} \, . 
\ee
This function obeys the elliptic transformation property~\eqref{elliptic} with index~$m$, but is not modular. However, it can 
be completed to a non-holomorphic function~$\wh \AA_{1,k}$ defined as~\cite{DMZ}:
\be \label{defFsmhat} 
  \wh \AA_{1,m}(\t,z) \= \AA_{1,m}(\t,z) \; + 
 2m \sum_{\text{\rm $\ell$ (mod $2m$)}} \vth^{(1)*}_{m,\ell}(\t)\,\vth_{m,\ell}(\t,z) 
\ee
that does transform like a \emph{holomorphic} Jacobi form of weight 1 and index $m$. 
Here, we have used, for $\,\ell\in\Z/2m\Z$, the standard theta function 
\be\label{deftheta1/2} 
  \vth_{m,\ell}(\t,z) \=  \sum_{\l\inn\Z \atop \l = \ell \mypmod{2m}} q^{\l^2/4m} \, \zeta^{\l},  
\ee
its first Taylor coefficient 
\be\label{deftheta3/2} 
  \vth^{(1)}_{m,\ell}(\t) \=  \left. \frac{1}{2\pi i} \frac{d}{dz} \vth_{m,\ell}(\t,z) \right|_{z=0} \= 
  \sum_{\l\inn\Z \atop \l = \ell \mypmod{2m}}\l\,q^{\l^2/4m} \, ,  
\ee
and its Eichler integral
\be\label{defEichler3/2} 
  \vth^{(1)*}_{m,\ell}(\t) \=   \sum_{\l\inn\Z \atop \l = \ell \mypmod{2m}}
      \sgn(\l)\,\,\erfc\bigl(2|\l|\sqrt{\pi m\t_2}\bigr)\,q^{-\l^2/4m}  \, .\ee
The function~$\wh \AA_{1,m}$ obeys the equation~\cite{DMZ} 
%CHECK normalization ZZZ
\be \label{holanomA1}
-\frac{2i}{\sqrt{k}} \t_{2}^{1/2} \p_{\tbar} \wh \AA_{1,m}(\t,z) \= \sum_{\ell \mypmod{2m}}\overline{\vth^{(1)}_{m,\ell}(\t)}\,\vth_{m,\ell}(\t,z) \, .
\ee
The above facts are summarized by saying that the functions $\AA_{1,m}(\t,z)$ can be completed to 
Jacobi forms $\wh \AA_{1,km}$ of weight 1 and index~$m$, with \emph{shadow} 
$\sum_{\ell\;\text{(mod $2m$)}}\overline{\vth^{(1)}_{m,\ell}(\t)}\,\vth_{m,\ell}(\t,z)$.

%\bibliography{GLSMCigar}{}

\begin{thebibliography}{10}

\bibitem{Schellekens:1986yi}
A.~Schellekens and N.~Warner, {\it {Anomalies and modular invariance in string
  theory}},  {\em Phys.Lett.} {\bf B177} (1986) 317.

\bibitem{Schellekens:1986xh}
A.~Schellekens and N.~Warner, {\it {Anomalies, Characters and Strings}},  {\em
  Nucl.Phys.} {\bf B287} (1987) 317.

\bibitem{Pilch:1986en}
K.~Pilch, A.~Schellekens, and N.~Warner, {\it {Path integral calculation of
  string anomalies}},  {\em Nucl.Phys.} {\bf B287} (1987) 362.

\bibitem{Witten:1986bf}
E.~Witten, {\it {Elliptic genera and quantum field theory}},  {\em Commun.
  Math. Phys.} {\bf 109} (1987) 525.

\bibitem{Witten:1987cg}
E.~Witten, {\it {The index of the Dirac operator in loop space}}, (1987) PUPT-1050.

\bibitem{Alvarez:1987wg}
O.~Alvarez, T.~P. Killingback, M.~L. Mangano, and P.~Windey, {\it {String
  theory and loop space index theorems}},  {\em Commun. Math. Phys.} {\bf 111}
  (1987) 1.

\bibitem{Alvarez:1987de}
O.~Alvarez, T.~P. Killingback, M.~L. Mangano, and P.~Windey, {\it {The
  Dirac-Ramond operator in string theory and loop space index theorems}}, .
  Invited talk presented at the Irvine Conf. on Non- Perturbative Methods in
  Physics, Irvine, Calif., Jan 5-9, 1987.

\bibitem{Eichler:1985ja}
M.~Eichler and D.~Zagier, {\em The Theory of Jacobi Forms}.
\newblock Birkh{\"a}user, 1985.

\bibitem{Zagier:Ellgen}
D.~Zagier, {\it {Note on the Landweber-Stong elliptic genus}},  {\em {Elliptic
  Curves and Modular Forms in Algebraic Topology, Proceedings, Princeton 1986,
  Lecture Notes 1326, Springer-Verlag}} (1988) 216--224.

\bibitem{Hori:2001ax}
K.~Hori and A.~Kapustin, {\it {Duality of the fermionic 2-D black hole and N=2
  liouville theory as mirror symmetry}},  {\em JHEP} {\bf 0108} (2001) 045,
  [\href{http://xxx.lanl.gov/abs/hep-th/0104202}{{\tt hep-th/0104202}}].

\bibitem{Troost:2010ud}
J.~Troost, {\it {The non-compact elliptic genus: mock or modular}},  {\em JHEP}
  {\bf 1006} (2010) 104, [\href{http://xxx.lanl.gov/abs/1004.3649}{{\tt
  arXiv:1004.3649}}].

\bibitem{Eguchi:2010cb}
T.~Eguchi and Y.~Sugawara, {\it {Non-holomorphic Modular Forms and SL(2,R)/U(1)
  Superconformal Field Theory}},  {\em JHEP} {\bf 1103} (2011) 107,
  [\href{http://xxx.lanl.gov/abs/1012.5721}{{\tt arXiv:1012.5721}}].

\bibitem{Ashok:2011cy}
S.~K. Ashok and J.~Troost, {\it {A Twisted Non-compact Elliptic Genus}},  {\em
  JHEP} {\bf 1103} (2011) 067, [\href{http://xxx.lanl.gov/abs/1101.1059}{{\tt
  arXiv:1101.1059}}].

\bibitem{Zwegers:2002}
S.~P. Zwegers, {\it {Mock theta functions}},  {\em {Thesis, Utrecht}} (2002).

\bibitem{Zagier:2007}
D.~Zagier, {\it {Ramanujan's mock theta functions and their applications
  [d'apr$\grave{\textrm{e}}$s Zwegers and Bringmann-Ono]}},  {\em
  {S{\'e}minaire BOURBAKI, 60 $\grave{e}me$ ann{\'e}e, 2006-2007}} {\bf {986}}
  (2007).

\bibitem{DMZ}
A.~Dabholkar, S.~Murthy, and D.~Zagier, {\it {Quantum Black Holes, Wall
  Crossing, and Mock Modular Forms}},
  \href{http://xxx.lanl.gov/abs/1208.4074}{{\tt arXiv:1208.4074}}.

\bibitem{Ashok:2013zka}
S.~K. Ashok and J.~Troost, {\it {Elliptic genera and real Jacobi forms}},
  \href{http://xxx.lanl.gov/abs/1310.2124}{{\tt arXiv:1310.2124}}.

\bibitem{Witten:1993yc}
E.~Witten, {\it {Phases of N=2 theories in two-dimensions}},  {\em Nucl.Phys.}
  {\bf B403} (1993) 159--222,
  [\href{http://xxx.lanl.gov/abs/hep-th/9301042}{{\tt hep-th/9301042}}].

\bibitem{Hori:2002cd}
K.~Hori and A.~Kapustin, {\it {World sheet descriptions of wrapped NS
  five-branes}},  {\em JHEP} {\bf 0211} (2002) 038,
  [\href{http://xxx.lanl.gov/abs/hep-th/0203147}{{\tt hep-th/0203147}}].

\bibitem{Benini:2013nda}
F.~Benini, R.~Eager, K.~Hori, and Y.~Tachikawa, {\it {Elliptic genera of
  two-dimensional N=2 gauge theories with rank-one gauge groups}},
  \href{http://xxx.lanl.gov/abs/1305.0533}{{\tt arXiv:1305.0533}}.

\bibitem{deWit:1985bn}
B.~de~Wit and M.~T. Grisaru, {\it {Compensating fields and anomalies}},  {\em
  {Batalin, I.A. (Ed.) et al.: Quantum field theory and quantum statistics}}
  {\bf 2} (1985) 411--432.


\bibitem{Ashok:2013pya}
Sujay~K.~Ashok, Nima Doroud, and Jan Troost, {\it {Localization and real Jacobi forms}},
  \href{http://xxx.lanl.gov/abs/1311.1110}{{\tt arXiv:1311.11103}}.


\bibitem{Witten:1991yr}
E.~Witten, {\it {On string theory and black holes}},  {\em Phys.Rev.} {\bf D44}
  (1991) 314--324.

\bibitem{Murthy:2003es}
S.~Murthy, {\it {Notes on noncritical superstrings in various dimensions}},
  {\em JHEP} {\bf 0311} (2003) 056,
  [\href{http://xxx.lanl.gov/abs/hep-th/0305197}{{\tt hep-th/0305197}}].

\bibitem{Konishi:1983hf}
K.~Konishi, {\it {Anomalous Supersymmetry Transformation of Some Composite
  Operators in SQCD}},  {\em Phys.Lett.} {\bf B135} (1984) 439.

\bibitem{Eguchi:2004yi}
T.~Eguchi and Y.~Sugawara, {\it {SL(2,R) / U(1) supercoset and elliptic genera
  of noncompact Calabi-Yau manifolds}},  {\em JHEP} {\bf 0405} (2004) 014,
  [\href{http://xxx.lanl.gov/abs/hep-th/0403193}{{\tt hep-th/0403193}}].

\bibitem{Gawedzki:1991yu}
K.~Gawedzki, {\it {Noncompact WZW conformal field theories}},
  \href{http://xxx.lanl.gov/abs/hep-th/9110076}{{\tt hep-th/9110076}}.

\bibitem{Gawedzki:1988nj}
K.~Gawedzki and A.~Kupiainen, {\it {Coset Construction from Functional
  Integrals}},  {\em Nucl.Phys.} {\bf B320} (1989) 625.

\bibitem{Karabali:1989dk}
D.~Karabali and H.~J. Schnitzer, {\it {BRST Quantization of the Gauged WZW
  Action and Coset Conformal Field Theories}},  {\em Nucl.Phys.} {\bf B329}
  (1990) 649.

\bibitem{Schnitzer:1988qj}
H.~J. Schnitzer, {\it {A Path Integral Construction of Superconformal Field
  Theories From a Gauged Supersymmetric {Wess-Zumino-Witten} Action}},  {\em
  Nucl.Phys.} {\bf B324} (1989) 412.

\bibitem{Ashok:2014nua}
S.~Ashok, E.~Dell'Aquila, and J.~Troost, {\it {Higher Poles and Crossing Phenomena from Twisted
Genera}},   
  [\href{http://xxx.lanl.gov/abs/1404.7396}{{\tt arXiv:1404.7396}}].

\bibitem{Pestun:2007rz}
V.~Pestun, {\it {Localization of gauge theory on a four-sphere and
  supersymmetric Wilson loops}},  {\em Commun.Math.Phys.} {\bf 313} (2012)
  71--129, [\href{http://xxx.lanl.gov/abs/0712.2824}{{\tt arXiv:0712.2824}}].

\bibitem{Gadde:2013dda}
A.~Gadde and S.~Gukov, {\it {2d Index and Surface operators}},
  \href{http://xxx.lanl.gov/abs/1305.0266}{{\tt arXiv:1305.0266}}.

\bibitem{Benini:2012ui}
F.~Benini and S.~Cremonesi, {\it {Partition functions of N=(2,2) gauge theories
  on $S^2$ and vortices}},  \href{http://xxx.lanl.gov/abs/1206.2356}{{\tt
  arXiv:1206.2356}}.

\bibitem{Doroud:2012xw}
N.~Doroud, J.~Gomis, B.~Le~Floch, and S.~Lee, {\it {Exact Results in D=2
  Supersymmetric Gauge Theories}},  {\em JHEP} {\bf 1305} (2013) 093,
  [\href{http://xxx.lanl.gov/abs/1206.2606}{{\tt arXiv:1206.2606}}].

\bibitem{Dabholkar:2010uh}
A.~Dabholkar, J.~Gomes, and S.~Murthy, {\it {Quantum black holes, localization
  and the topological string}},  {\em JHEP} {\bf 1106} (2011) 019,
  [\href{http://xxx.lanl.gov/abs/1012.0265}{{\tt arXiv:1012.0265}}].

\bibitem{Dabholkar:2011ec}
A.~Dabholkar, J.~Gomes, and S.~Murthy, {\it {Localization $\&$ Exact
  Holography}},  {\em JHEP} {\bf 1304} (2013) 062,
  [\href{http://xxx.lanl.gov/abs/1111.1161}{{\tt arXiv:1111.1161}}].

\bibitem{Kraus:2006nb}
P.~Kraus and F.~Larsen, {\it {Partition functions and elliptic genera from supergravity}}, 
{\em JHEP} {\bf 0701} (2007) 002,
  [\href{http://xxx.lanl.gov/abs/hep-th/0607138}{{\tt hep-th/0607138}}].

\bibitem{Jockers:2012dk}
H.~Jockers, V.~Kumar, J.~M. Lapan, D.~R. Morrison, and M.~Romo, {\it
  {Two-Sphere Partition Functions and Gromov-Witten Invariants}},
  \href{http://xxx.lanl.gov/abs/1208.6244}{{\tt arXiv:1208.6244}}.

\bibitem{Gomis:2012wy}
J.~Gomis and S.~Lee, {\it {Exact Kahler Potential from Gauge Theory and Mirror
  Symmetry}},  {\em JHEP} {\bf 1304} (2013) 019,
  [\href{http://xxx.lanl.gov/abs/1210.6022}{{\tt arXiv:1210.6022}}].

\bibitem{Murthy:2013xpa}
S.~Murthy and V.~Reys, {\it {Quantum black hole entropy and the holomorphic
  prepotential of N=2 supergravity}},
  \href{http://xxx.lanl.gov/abs/1306.3796}{{\tt arXiv:1306.3796}}.

\bibitem{Harvey:2013mda}
J.~A. Harvey and S.~Murthy, {\it {Moonshine in Fivebrane Spacetimes}},
  \href{http://xxx.lanl.gov/abs/1307.7717}{{\tt arXiv:1307.7717}}.

\bibitem{Ginsparg:1988ui}
P.~H. Ginsparg, {\it {Applied conformal field theory}},
  \href{http://xxx.lanl.gov/abs/hep-th/9108028}{{\tt hep-th/9108028}}.

\bibitem{AlvarezGaume:1986es}
L.~Alvarez-Gaume, G.~W. Moore, and C.~Vafa, {\it {Theta Functions, Modular
  Invariance and Strings}},  {\em Commun.Math.Phys.} {\bf 106} (1986) 1--40.

\bibitem{Ashok:2012qy}
S.~K. Ashok and J.~Troost, {\it {Elliptic Genera of Non-compact Gepner Models
  and Mirror Symmetry}},  {\em JHEP} {\bf 1207} (2012) 005,
  [\href{http://xxx.lanl.gov/abs/1204.3802}{{\tt arXiv:1204.3802}}].

\bibitem{Ashok:2013kk}
S.~K. Ashok, S.~Nampuri, and J.~Troost, {\it {Counting Strings, Wound and
  Bound}},  {\em JHEP} {\bf 1304} (2013) 096,
  [\href{http://xxx.lanl.gov/abs/1302.1045}{{\tt arXiv:1302.1045}}].

\bibitem{Fendley:1992dm}
P.~Fendley and K.~A. Intriligator, {\it {Scattering and thermodynamics in
  integrable N=2 theories}},  {\em Nucl.Phys.} {\bf B380} (1992) 265--292,
  [\href{http://xxx.lanl.gov/abs/hep-th/9202011}{{\tt hep-th/9202011}}].

\bibitem{Aganagic:2001uw}
M.~Aganagic, K.~Hori, A.~Karch, and D.~Tong, {\it {Mirror symmetry in
  (2+1)-dimensions and (1+1)-dimensions}},  {\em JHEP} {\bf 0107} (2001) 022,
  [\href{http://xxx.lanl.gov/abs/hep-th/0105075}{{\tt hep-th/0105075}}].

\bibitem{McGreevy:2003dn}
J.~McGreevy, S.~Murthy, and H.~L. Verlinde, {\it {Two-dimensional superstrings
  and the supersymmetric matrix model}},  {\em JHEP} {\bf 0404} (2004) 015,
  [\href{http://xxx.lanl.gov/abs/hep-th/0308105}{{\tt hep-th/0308105}}].

\bibitem{Eguchi:2003ik}
T.~Eguchi and Y.~Sugawara, {\it {Modular bootstrap for boundary N = 2 Liouville
  theory}},  {\em JHEP} {\bf 0401} (2004) 025,
  [\href{http://xxx.lanl.gov/abs/hep-th/0311141}{{\tt hep-th/0311141}}].

\bibitem{Benini:2013xpa}
F.~Benini, R.~Eager, K.~Hori, and Y.~Tachikawa, {\it {Elliptic genera of 2d N=2
  gauge theories}},  \href{http://xxx.lanl.gov/abs/1308.4896}{{\tt
  arXiv:1308.4896}}.

\bibitem{Gritsenko:1999fk}
V.~Gritsenko, {\it {Elliptic genus of Calabi-Yau manifolds and Jacobi and
  Siegel modular forms}},  \href{http://xxx.lanl.gov/abs/math/9906190}{{\tt
  math/9906190}}.

\bibitem{Ooguri:1995wj}
H.~Ooguri and C.~Vafa, {\it {Two-dimensional black hole and singularities of CY
  manifolds}},  {\em Nucl.Phys.} {\bf B463} (1996) 55--72,
  [\href{http://xxx.lanl.gov/abs/hep-th/9511164}{{\tt hep-th/9511164}}].

\bibitem{Giveon:1999zm}
A.~Giveon, D.~Kutasov, and O.~Pelc, {\it {Holography for noncritical
  superstrings}},  {\em JHEP} {\bf 9910} (1999) 035,
  [\href{http://xxx.lanl.gov/abs/hep-th/9907178}{{\tt hep-th/9907178}}].

\bibitem{Giveon:1999px}
A.~Giveon and D.~Kutasov, {\it {Little string theory in a double scaling
  limit}},  {\em JHEP} {\bf 9910} (1999) 034,
  [\href{http://xxx.lanl.gov/abs/hep-th/9909110}{{\tt hep-th/9909110}}].

\end{thebibliography}
\bibliographystyle{JHEP}

\providecommand{\href}[2]{#2}\begingroup\raggedright\endgroup

\end{document}